\documentclass{aa}  
\usepackage{graphicx}
\usepackage{txfonts}
 \usepackage{natbib}
 \usepackage{comment}
\begin{document}
\title{Detectability of biosignatures \\ in warm, water-rich atmospheres}



   \author{B. Taysum,
          \inst{1}
           I. van Zelst,\inst{1,2}
          J. L. Grenfell\inst{1},
          F. Schreier,\inst{3}
          J. Cabrera\inst{1}
          and
          H. Rauer\inst{1,4}
          }

   \institute{ Institut für Planetenforschung (PF), Deutsches Zentrum für Luft- und Raumfahrt (DLR),, Rutherfordstraße 2, 12489 Berlin, Germany, \email{benjamin.taysum@dlr.de}
         \and
            Zentrum für Astronomie und Astrophysik (ZAA), Technische Universität Berlin (TUB), Hardenbergstraße 36, 10623, Berlin, Germany
            \and
            Institut für Methodik der Fernerkundung (IMF), German Aerospace Center (DLR), 82234 Wessling, Germany
        \and 
         Institut für Geologische Wissenschaften, Freie Universität Berlin (FUB), Malteserstraße 74-100, 12249, Berlin, Germany
             }
    \date{}
 
  \abstract
   {Warm rocky exoplanets within the habitable zone of Sun-like stars are favoured targets for current and future missions. Theory indicates these planets could be wet at formation and remain habitable long enough for life to develop. However, it is unclear to what extent an early ocean on such worlds could influence the response of potential biosignatures. 
   }
   {In this work we test the climate-chemistry response, maintenance, and detectability of biosignatures in warm, water-rich atmospheres with Earth biomass fluxes within the framework of the planned LIFE mission.
   }
   {We used the coupled climate-chemistry column model 1D-TERRA to simulate the composition of planetary atmospheres at different distances from the Sun, assuming Earth's planetary parameters and evolution. We increased the incoming instellation by up to 50 percent in steps of 10 percent, corresponding to orbits of 1.00 to 0.82 AU. Simulations were performed with and without modern Earth’s biomass fluxes at the surface. Theoretical emission spectra of all simulations were produced using the GARLIC radiative transfer model. LIFEsim was then used to add noise to and simulate observations of these spectra to assess how biotic and abiotic atmospheres of Earth-like planets can be distinguished.
   }
   {Increasing instellation leads to surface water vapour pressures rising from 0.01 bar (1.31\%, S = 1.0) to 0.61 bar (34.72\%, S = 1.5). In the biotic scenarios, the ozone layer survives because hydrogen oxide reactions with nitrogen oxides prevent the net ozone chemical sink from increasing. Methane is strongly reduced for instellations that are 20\% higher than that of the Earth due to the increased hydrogen oxide abundances and UV fluxes. Synthetic observations with LIFEsim, assuming a 2.0 m aperture and resolving power of a R = 50, show that ozone signatures at 9.6 $\mu$m reliably point to Earth-like biosphere surface fluxes of O$_2$ only for systems within 10 parsecs. The differences in atmospheric temperature structures due to differing H$_2$O profiles also enable observations at 15.0 $\mu$m to reliably identify planets with a CH$_4$ surface flux equal to that of Earth's biosphere. Increasing the aperture to 3.5 m  increases this range to 22.5 pc. } 
    {}

   \titlerunning{Detectability of biosignatures in warm, water-rich atmospheres}
   \authorrunning{Taysum et al.}
   \maketitle

\section{Introduction}
\label{sec:introduction} Rocky exoplanets in the inner habitable zone (IHZ) are at the cutting edge of exoplanetary research. Ongoing missions such as the Transiting Exoplanet Survey Satellite (TESS, \cite{ricker2010transiting}) are in the midst of discovering rocky planets in the so-called Venus Zone \citep{ostberg2023}. These warm worlds have rather small orbital periods and will be discovered and characterised before their cooler counterparts in the middle and outer habitable zone. Furthermore, recently adopted missions to Venus by ESA (EnVision,\cite{ghail_2016}) and NASA (Veritas and DAVINCI; \cite{smrekar_2020}, \cite{garvin2022}) will also have a knock-on effect on exo-Venus research.
Regarding volatile delivery, current theories suggest that rocky worlds forming in and around the IHZ of Sun-like stars will likely accrete water inventories that are broadly comparable with objects formed in the mid habitable zone  \citep[see e.g.][]{lammer2018}. Recent studies \citep[e.g.][]{way_2020,honing_2021} suggest an extended early habitability of Venus and Venus-like worlds \citep[see e.g.][]{Westall2023}. The evolution of Venus-like planets and possible early habitability is a central issue in exoplanet science.

Current theory does not rule out modest, warm, water-rich atmospheres due to ocean evaporation as warm, rocky planets evolve through and beyond their habitable phase \citep{kasting1993earth,zahnle2010earth,ElkinsTanton2012}. During early evolution, however, rocky exoplanets can feature giant, hot steam atmospheres, typically from the close of the magma ocean (MO) period when crustal formation leads to strong volatile outgassing. These steam atmospheres are believed to play a crucial role in the early evolution of terrestrial planets and can eventually condense to form global oceans and  hence lead to long-lived habitable conditions.  Early Earth likely featured such a giant steam atmosphere at the close of its final MO period after the Moon-forming event
\citep{ElkinsTanton2008magmasolid,lebrun2013thermal,sossi2020redox,gaillard2022redox,sossi2023solubility,Dorn2021}.

Regarding Venus, \citet{ORourke2023} recently reviewed planetary evolution and note that early climate runaway, and hence the timing and extent of a potential giant steam atmosphere phase, is likely sensitive to the global coverage and refractive properties of its clouds, which are not well known. For example, \citet{Salvador2023} reviewed water delivery and the blanket effect influencing the giant steam atmosphere climate on early Venus. \citet{Westall2023} reviewed Venus's evolution, including volatile cycling and the consequences for long-term habitability. \citet{Lichtenegger2016} suggested that several hundred bars of steam atmosphere on early Venus could have been removed in the first 100 Myr, mainly due to non-thermal escape, depending on the uncertain extreme-UV flux of the early Sun. \citet{Way2020} discuss the implications of the deuterium ($^2$H) and protium ($^1$H) ratio (D/H) data from Pioneer Venus, which suggest an extended shallow ocean on early Venus. The 3D model study by \citet{Turbet2021}, however, suggests that the early Venusian giant steam atmosphere did not condense. Resolving these issues requires improved (more accurate and spatially extended) isotope and other proxy data, as discussed in, for example, \citet{Widemann2023}. 
 
Regarding rocky exoplanets, giant steam atmospheres are also likely during the early evolution. Their duration will depend on the interplay between, for example, outgassing, escape, and climate (which are not well determined) and could vary from several thousand to several million years after the close of the MO phase, as discussed in \citet{nikolaou2019factors}. 
Numerous studies have addressed the emergence, evolution, and climate of early, giant steam atmospheres by applying radiative-convective models. \cite{abe1985formation} highlight the effectiveness of a steam-dominated atmosphere in thermally insulating the surface, but their assumption of a grey atmosphere
(i.e. the absorption coefficients are assumed to be constant) led to an underestimation of surface warming. \citet{kasting1988runaway} applied a 1D climate model to study the response of Earth-like atmospheres to large increases in instellation. More recently, \citet{marcq2017thermal} used a 1D radiative-convective model of H$_2$O-CO$_2$ atmospheres to study the thermal blanketing effect of steam atmospheres.  \citet{katyal2019evolution,katyal2020effect} used a radiative-convective model to study the evolution of a steam atmosphere around the time of Earth's final MO.
\citet{Kite2021} suggested that inward migrating sub-Neptunes can evolve into hot, rocky exoplanets and attain thick steam atmospheres from their original gas mantles. \citet{Harman2022} suggest that a 100 bar steam atmosphere on TOI-1266c could be detected with about 20 hours of observing time by the \textit{James Webb} Space Telescope. 

Earth-like atmospheric biosignatures and their responses to varying planetary and stellar parameters have been widely studied, for example: the response to the central star (e.g. \citealt{Segura2005,Grenfell2007}), the response over planetary evolution timescales (e.g. \citealt{Gebauer2018,Rugheimer2018}), and the response to clouds (e.g. \citealt{Kitzmann2011}), to name but a few. None of these studies, however, investigated atmospheric biosignatures with consistent climate chemistry throughout the IHZ for conditions in which warm, water-rich atmospheres can form. Therefore, despite a rich, emerging literature focusing on, for example, early steam atmospheres in the Solar System and beyond, to our knowledge there are no dedicated climate-chemistry coupled studies for warm rocky exoplanets with water-rich atmospheres from evaporating oceans. Such models require flexible radiative transfer modules that can operate over a large range  of atmospheres as well as comprehensive chemistry modules. Including coupled climate-chemistry models is important for three main reasons: (1) The gas-phase (photo)chemistry of water-rich atmospheres could directly affect atmospheric biosignatures; water-rich atmospheres can generate strong abundances of hydrogen oxides, HO$_\mathrm{X}$ (=OH+HO$_2$), via the photolytic breakdown of water by UV radiation. HO$_\mathrm{X}$ can in turn destroy potential biosignatures such as ozone via catalytic cycles and methane via direct gas-phase reactions with the hydroxyl (OH) radical, which can also stabilise climate gases such as CO$_2$ \citep{Yung1998-wk}. (2) Via abiotic processes, steam can rapidly photolyse in the upper atmosphere. The resulting H atoms can be lost via escape, whereas the ground-state oxygen (O($^3$P)) atom is left behind. These oxygen atoms can self-react, leading to abiotic production of molecular oxygen and ozone (see e.g. \citealt{Meadows2017}).
(3) Due to UV shielding and climate blanketing, steam atmospheres shield UV and heat, which can directly affect photolysis and temperature-dependent gas-phase chemical reactions.

In this work we address these issues by performing coupled climate-chemistry model simulations of warm rocky planets with oceans for scenarios with and without Earth-like biospheres at the surface across the IHZ.\ These simulations produce high abundances ($>$ 10\%) of water vapour at higher instellation runs. 

\section{Model and methods}\label{sec:methods}

\begin{table}[ht]
\caption{Biomass surface fluxes, dry deposition velocities, and volcanic outgassing fluxes  for trace gas  species used in all model runs in this work.}
\label{tab:lower_boundary_flux}
\centering
\resizebox{\linewidth}{!}{
\begin{tabular}{cccc}
Species    & \begin{tabular}[c]{@{}c@{}}Surface Flux\\ (molec. cm$^{-2}$ s$^{-1}$)\end{tabular} & \begin{tabular}[c]{@{}c@{}}Deposition\\ Velocity\\ (cm s$^{-1}$)\end{tabular} & \begin{tabular}[c]{@{}c@{}}Volcanic Flux\\ (molec. cm$^{-2}$ s$^{-1}$)\end{tabular} \\ \hline
CO$_2$     & -2.92$\times$10$^{11}$                                                             & -                                                                             & 3.18$\times$10$^{10}$                                                               \\
O$_2$      & 1.18$\times$10$^{12}$                                                              & 1.00$\times$10$^{-8}$                                                         & -                                                                                   \\
O$_3$      & -                                                                                  & 0.40                                                                          & -                                                                                   \\
H$_2$      & 9.29$\times$10$^{10}$                                                              & 2.00$\times$10$^{-2}$                                                         & 3.75$\times$10$^{9}$                                                                \\
CO         & 2.23$\times$10$^{11}$                                                              & 3.00$\times$10$^{-2}$                                                         & 3.74$\times$10$^{8}$                                                                \\
CH$_4$     & 1.40$\times$10$^{11}$                                                              & 1.55$\times$10$^{-4}$                                                         & 1.12$\times$10$^{8}$                                                                \\
C$_2$H$_2$ & 9.48$\times$10$^{8}$                                                               & 2.00$\times$10$^{-2}$                                                         & -                                                                                   \\
C$_2$H$_6$ & 1.56$\times$10$^{9}$                                                               & 2.00$\times$10$^{-2}$                                                         & 5.10$\times$10$^{6}$                                                                \\
C$_3$H$_8$ & 1.50$\times$10$^{9}$                                                               & 2.00$\times$10$^{-2}$                                                         & 2.29$\times$10$^{6}$                                                                \\
N$_2$O     & 1.44$\times$10$^{9}$                                                               & 1.00$\times$10$^{-5}$                                                         & -                                                                                   \\
NO         & 2.80$\times$10$^{9}$                                                               & 1.60$\times$10$^{-2}$                                                         & -                                                                                   \\
H$_2$S     & 1.86$\times$10$^{9}$                                                               & 1.50$\times$10$^{-2}$                                                         & 1.89$\times$10$^{9}$                                                                \\
SO$_2$     & 1.70$\times$10$^{10}$                                                              & 1.00                                                                          & 1.34$\times$10$^{10}$                                                               \\
NH$_3$     & 4.40$\times$10$^{9}$                                                               & 1.71                                                                          & -                                                                                   \\
OCS        & 1.84$\times$10$^{8}$                                                               & 1.00$\times$10$^{-2}$                                                         & 2.67$\times$10$^{6}$                                                                \\
HCN        & 1.44$\times$10$^{8}$                                                               & 3.15$\times$10$^{-2}$                                                         & -                                                                                   \\
HNO$_3$    & -                                                                                  & 4.00                                                                          & -                                                                                   \\
CH$_3$OH   & 3.64$\times$10$^{10}$                                                              & 1.26                                                                          & -                                                                                   \\
CS$_2$     & 6.14$\times$10$^{8}$                                                               & 2.00$\times$10$^{-2}$                                                         & 6.23$\times$10$^{6}$                                                                \\
CH$_3$Cl   & 2.19$\times$10$^{8}$                                                               & 1.50$\times$10$^{-3}$                                                         & -                                                                                   \\
HCl        & 6.45$\times$10$^{9}$                                                               & 0.80                                                                          & 4.42$\times$10$^{8}$                                                                                                                                        
\end{tabular}
}
\end{table}

\subsection{Atmospheric column model, 1D-TERRA}

1D-TERRA is a global-mean, stationary, coupled convective-climate-photochemical model consisting of one 
hundred layers extending from the planetary surface to 10$^{-5}$ hPa. The convective-climate module 
\citep{scheucher2020} applies an adiabatic lapse rate in the lower atmosphere,  with a convective regime and 
radiative transport above according to the Schwarzschild criterium. 
The model is valid for pressures up to 1000 bar and temperatures up to 1000 K and can reproduce the 
atmospheric parameters and compositions of modern Earth, Venus, and Mars \citep{scheucher2020}. 
The climate module  uses the random overlap method for the frequency integration from 100,000
cm$^{-1}$ (100 nm) to 0 cm$^{-1}$ and the delta two-stream approximation for the angular integration. 
Up to 20 absorbing species in the visible/IR and 81 cross-sections in the UV-visible are 
available. The model includes Rayleigh scattering and the option to include self and foreign continua. 
Convective adjustment is considered applying the dry adiabat or the wet adiabat for H$_2$O and CO$_2$ 
condensibles and including adjustable ocean reservoirs \citep{manabe1967thermal}. 

The chemistry module from \cite{wunderlich2020distinguishing} features 1127 reactions for 128 species 
including photolysis for 81 absorbers. Dry and wet deposition as well as biomass, lightning
\citep{NOxProductioninLightning} and 
volcanic mass fluxes can be included flexibly. Adaptive eddy mixing depending on concentration and 
temperature profiles is employed, using the approach similar to \cite{gao2015stability} 
that uses the equations developed by \cite{Gierasch1985}.
\subsection{1D-TERRA boundary parameters}

Table \ref{tab:lower_boundary_flux} lists the surface fluxes, volcanic
outgassing fluxes, and dry deposition velocities for trace gas 
species used in all model runs in this work. Surface fluxes deliver 
the species to the first model layer (centred at $\sim$ 500 m) only, whereas volcanic fluxes are 
evenly distributed across the lower 10 km of the model domain. For all species
except CO$_2$ and O$_2$, the surface fluxes are the sum of the modern 
Earth's biogenic and anthropogenic emissions \citep{wunderlich2020distinguishing}. 
Volcanic outgassing fluxes are similarly representative of the modern Earth 
\citep{catling2017atmospheric,khalil1984global,etiope2009earth,pyle2009halogens}.

The values for CO$_2$ and O$_2$ surface fluxes were chosen such that the model run 
with a solar constant value of 1.0 approximately reproduces the modern Earth atmospheric CO$_2$ and O$_2$
abundances (355 ppm and 0.21 mol/mol, respectively). The negative flux used for CO$_2$ 
represents surface uptake processes not included in our model involving, for example, biological and sub-surface processes.

1D-TERRA uses the Manabe-Wetherald relative
humidity profile \citep{manabe1967thermal} in
all simulations, which recreates the mean 
atmospheric water vapour profile of the modern
Earth for a solar constant of 1.0 with a fixed surface relative humidity of
80\%. The
\cite{NOxProductioninLightning} lightning model was used to produce modern-Earth concentrations
of NO and NO$_2$ within the troposphere,
and rainout rates of trace gas species were accounted for using the
\cite{giorgi1985rainout} approach with effective Henry's law 
constants via temperature-dependent parameterisations
from \cite{sander2015compilation}.  

\subsection{GARLIC spectral model}

GARLIC (General purpose Atmospheric Radiative transfer Line-by-line Infrared-microwave Code) was used to calculate theoretical transmission and emission spectra based on input of temperature and composition from 1D-TERRA output.
GARLIC uses an optimised Voigt function with a three-grid approach for line-by-line modelling of molecular cross-sections using HITRAN or GEISA data as well as continua, collision induced absorption, and Rayleigh and aerosol extinction.
The Jacobians (derivatives with respect to the unknowns of the atmospheric inverse problem) are constructed via automatic differentiation \citep{Schreier15}. GARLIC has been validated via intercomparison with other codes \citep{Schreier18agk} and by comparing its synthetic spectra with observed spectra for modern Earth and Venus \citep{Schreier18ace}. Further details can be found in \citet{Schreier2014} and \cite{Schreier19p}.

\subsection{Scenarios}

We modelled six scenarios with 1D-TERRA, simulating rocky exoplanets assuming Earth's development and biomass across the IHZ (1.00 AU; scenario 1, $S = 1.0$) with instellation increasing at intervals of 0.1, resulting in $S = 1.1$ (0.95 AU, scenario 2), $S = 1.2$ (0.91 AU, scenario 3), $S = 1.3$ (0.88 AU, scenario 4), $S = 1.4$ (0.85 AU, scenario 5), and $S = 1.5$ (0.82 AU, scenario 6). To assess the impact of the active biosphere on the atmospheric biomarkers, 
the runs were repeated with all positive surface fluxes in Table 
\ref{tab:lower_boundary_flux} removed. The negative surface flux of CO$_2$ was re-tuned to approximately reproduce
the CO$_2$ volume mixing ratio (VMR) of modern Earth for the abiotic $S$ = 1.0 case. With less HO$_\mathrm{X}$ in the mesosphere (pressures below approximately 0.01 hPa)
from lower stratospheric H$_2$O (discussed in Sect. 3.4), recycling of CO to CO$_2$ via
OH reactions is less efficient, and the lower CH$_4$ content 
from surface biology lowers the rate of CO$_2$ production
from CH$_4$ oxidation. An additional caveat is that
without the uptake of CO$_2$ from biological activity (e.g. photolysis,
soil uptake)
at the surface, abiotic planets with equal CO$_2$ outgassing
fluxes will result in greater CO$_2$ atmospheric abundances
than their biotic counterparts \citep{haqq2016limit}. However, 
to focus on the atmospheric chemistry and responses 
appropriately, we chose
to study biotic and abiotic worlds that posses the same
atmospheric CO$_2$ abundances. To achieve this, we lowered the CO$_2$ surface flux  to 
-2.286$\times$10$^{9}$ molecules cm$^{-2}$ s$^{-1}$ for all the abiotic runs. 
\begin{table}[h!]
\centering
\label{tab:surftemps}
\caption{Solar constants studied in this work for the six biotic model scenarios, with corresponding distances to the Sun, the resultant amount of H$_2$O at the model surface in percentage by volume and partial pressure, and the surface temperature.}
\begin{tabular}{cccc}
\hline\hline                 
\begin{tabular}[c]{@{}c@{}}Solar\\ Constant\end{tabular} & \begin{tabular}[c]{@{}c@{}}Distance\\ from Sun\\ (AU)\end{tabular} & \begin{tabular}[c]{@{}c@{}}H$_2$O \\ at Surface\\ (\% / bars)\end{tabular} & \begin{tabular}[c]{@{}c@{}}Surface\\ Temperature\\ (K) \end{tabular} \\
\hline               
1.0                                                      & 1.00                                                               & 1.31 / 0.01                                                                & 287.70\\
1.1                                                      & 0.95                                                               & 3.51 / 0.04                                                                & 304.41\\
1.2                                                      & 0.91                                                               & 7.29 / 0.08                                                                & 318.93\\
1.3                                                      & 0.88                                                               & 13.83 / 0.17                                                               & 334.07\\
1.4                                                      & 0.85                                                               & 22.62 / 0.32                                                               & 348.69 \\
1.5                                                      & 0.82                                                               & 34.72 / 0.61                                                               & 365.42\\
\hline   
\end{tabular}
\label{tab:Scenarios}
\end{table}

\subsection{LIFEsim astrophysical noise generator}

LIFEsim, developed by \cite{dannert2022lifesim} and used in exoplanet retrieval studies such as \cite{quanz2022large}, \cite{konrad2022large}, and \cite{alei2022earthanalog}, was used to
add astrophysical noise signals to the emission
spectra produced by the GARLIC radiative transfer code. The noise sources included are photon noise
from stellar leakage, exozodi disks, the planet
emission, and the local exozodi disk. 
Details of these noise sources are presented
in \cite{dannert2022lifesim} and 
\cite{quanz2022large}. There is expected to be significantly less instrument noise than photon shot noise in the present version of LIFEsim, and it does not take instrument noise into consideration 

The same simulation parameters used in 
\cite{alei2022earthanalog} are adopted in this
manuscript, referred to as the `baseline' LIFE
scenario. This configuration is expected to be capable
of measuring the emission spectra of Earth at
a distance of 10 pc with a wavelength-integrated
signal to noise ratio of 9.7 for a total 
measurement time of approximately 2.3 days 
\citep{dannert2022lifesim}.
The quantum efficiency of the interferometer's detector is equal to 0.7, 
the wavelength range is set to 5.0--18.5 $\mu$m,
the spectral resolution is R = 50, aperture 
diameter d = 2.0 m, and the exozodi level is set
to 3 times that of the local zodiacal dust.
LIFEsim is ran here across a grid of integration times (24, 48,
120, 240 hours) and distances from the Solar System (5--30 parsecs in intervals
of 5 pc).

The statistical significance of the biotic signal at each wavelength band was calculated by dividing the absolute difference in retrieved flux for
the biotic and abiotic scenarios by the noise calculated via
LIFEsim ($\sigma_{Bio}$):

\begin{equation*}
    \mathrm{Significance}(\lambda) = \frac{|F_{Bio} - F_{Abio}|(\lambda)}{\sigma_{Bio}(\lambda)} .
\end{equation*}

\section{Results}\label{sec:results}

   \begin{figure*}[h]
   \centering
   \includegraphics[width=\textwidth]{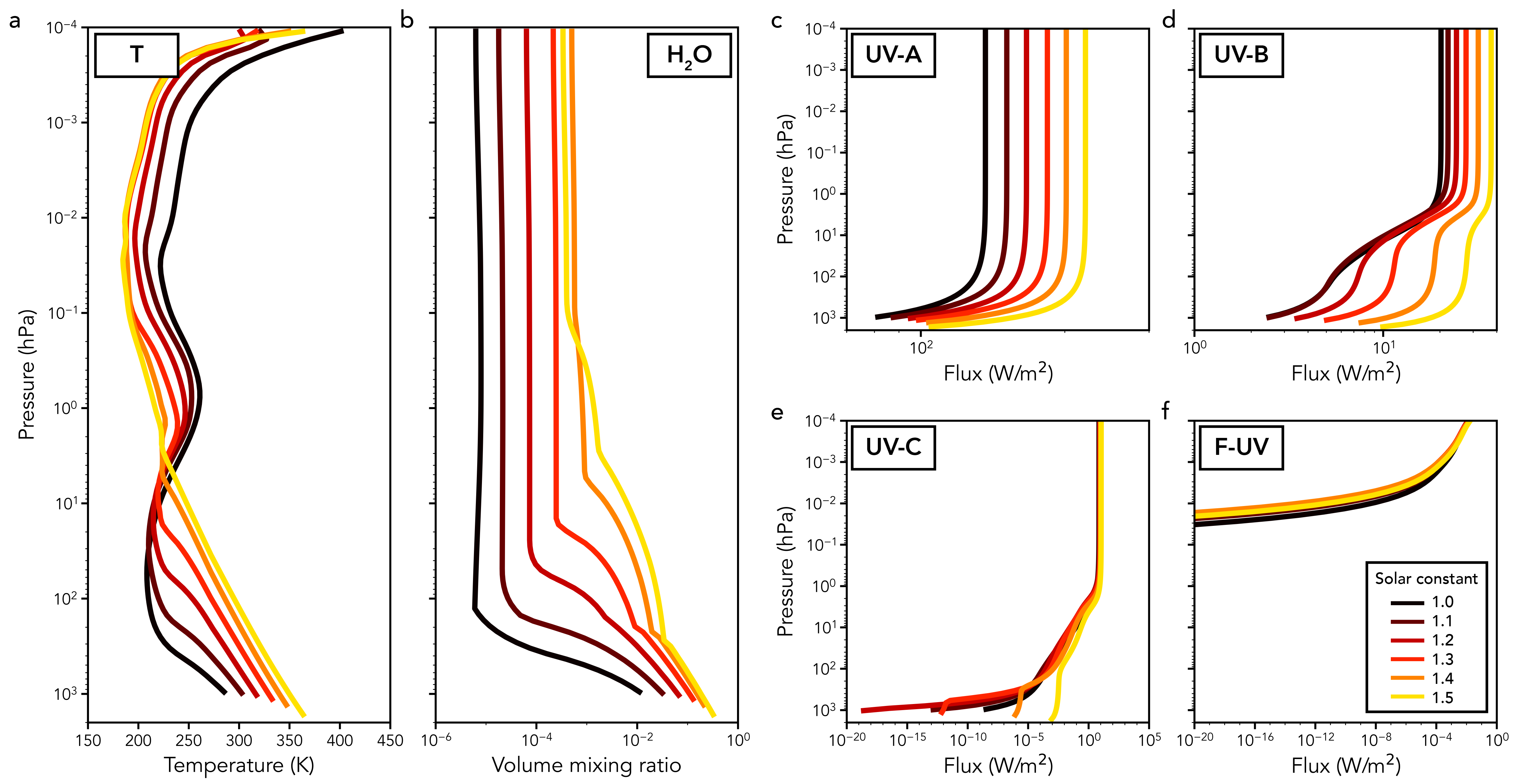}
   \caption{Temperature (a), H$_2$O vapour volume mixing ratio in mol/mol (b), and the UV-A (c), UV-B (d), UV-C (e), and FUV 
   (f) fluxes calculated by 1D-TERRA for a (biotic) Earth-like planet with the mean incoming instellation increased from 1.0 (modern Earth) up by a factor of 1.5 (orbiting a Sun-like star at 0.82 AU). }
         \label{fig:climate}
    \end{figure*}

Figure \ref{fig:climate} shows 1D-TERRA output of atmospheric temperature (Kelvin, panel a) and water
vapour (VMR; panel b) profiles, alongside the UV-A ($\lambda$ = 317.5 -- 405 nm), UV-B ($\lambda$ = 281.7 -- 317.5 nm), 
UV-C ($\lambda$ = 175.0 -- 281.7 nm), and far-UV ($\lambda$ = 100.0 -- 175.0 nm) radiation fluxes (W m$^{-2}$; panels c to f, respectively).
Regarding temperature,
the stronger instellation leads to an increase in surface temperature from 287.7~K (scenario 1) to 365.4~K (scenario 6). The cold trap moves upwards due to expansion of the atmospheric column in the troposphere and becomes weaker. The adiabatic lapse rate in the lower atmosphere changes gradient because warmer temperatures  lead to ocean evaporation,  which modifies atmospheric composition and heat capacity.
On increasing the instellation, firstly the temperature inversion (the rate of change of temperature with increasing altitude) is weakened. The inversion serves as a barrier to upward mixing, which acts as a `lid' on the troposphere, preventing mixing of water upwards into the stratosphere; this barrier effect is therefore weakened and results in 
greater H$_2$O quantities reaching the stratosphere. Secondly, the minimum temperature at the inversion increases from about 210K (scenario 1) to about 250K (scenario 6). This also suggests a weakening of the cold trap since it is less able to freeze out tropospheric water. The surface warming is accompanied by cooling in the middle atmosphere. This arises due to an enhanced classical greenhouse driven by increased ocean evaporation. Increasing H$_2$O 
abundances within the troposphere trap heat with greater efficiency, decreasing the upward flux of long-wave 
radiation to altitudes above the tropopause whilst causing surface temperatures to rise. The lower flux of IR
radiation entering the stratosphere that travels upwards from the troposphere causes regions with pressure below
10 hPa to cool, and the inversion layer to disappear by $S = 1.4$.

Regarding water vapour, the surface water VMRs range from 0.013 (P$_\mathrm{H2O}$ = 0.01 bar) for $S = 1.0$ to 0.347 (P$_\mathrm{H2O}$ = 0.6 bar) for $S = 1.5$ at the model surface. The increasing surface temperatures elevate 
the water vapour saturation pressures within the 
troposphere. As a fixed relative humidity of 80\%
is adopted within the troposphere, the resulting
partial pressure of water vapour increases within the 
model. This approximates the effects expected from the
evaporation of the Earth's surface oceans as the 
surface temperatures progressively increase.
At $S = 1.5$, a total surface pressure of about 1.6 bar occurs since we also assumed one bar of combined N$_2$ and O$_2$ for the modern Earth. On increasing instellation, the transition region from tropospheric water (with its adiabatic lapse rate) to stratospheric water (mainly affected by upward mixing and chemical production from methane oxidation) moves upwards due to expansion and results in a convex-shaped bulge for the higher instellations, associated with, for example, increasing upward water transport and due to the imposed Manabe-Wetherald relative humidity isoprofile.

UV-A radiation in Fig. \ref{fig:climate} penetrates efficiently through the atmosphere down to about 100 hPA, below which higher pressures lead to its exponential absorption in all scenarios.
UV-B is mainly absorbed by O$_3$.
The presence
of Earth's O$_3$ layer centred at about 10 hPa (around 30km on modern Earth) effectively shields the surface from the incoming UV-B flux,
resulting in the absorption `shoulder' slope seen in Fig. \ref{fig:climate}d.
In the free troposphere the flux gradient of UV-B increases more strongly with altitude for the higher instellation scenarios. This is likely related to the ozone response, as discussed in the next section.
UV-C radiation decreases rapidly for scenarios $S = 1.0$ to $S = 1.2$ for pressures greater than about 100 hPa, mainly due to increasing water vapour and atmospheric pressure. For scenarios $S = 1.3$ to $S =1.5,$ however, the radiation is more efficient at penetrating down to the surface. This is related to the ozone response as discussed in the next section. Far-UV radiation is mainly absorbed in the mesosphere via water vapour for all scenarios.

\subsection{Climate-chemistry response of biosignatures}

\begin{figure}
\centering
    \includegraphics[width=\hsize]{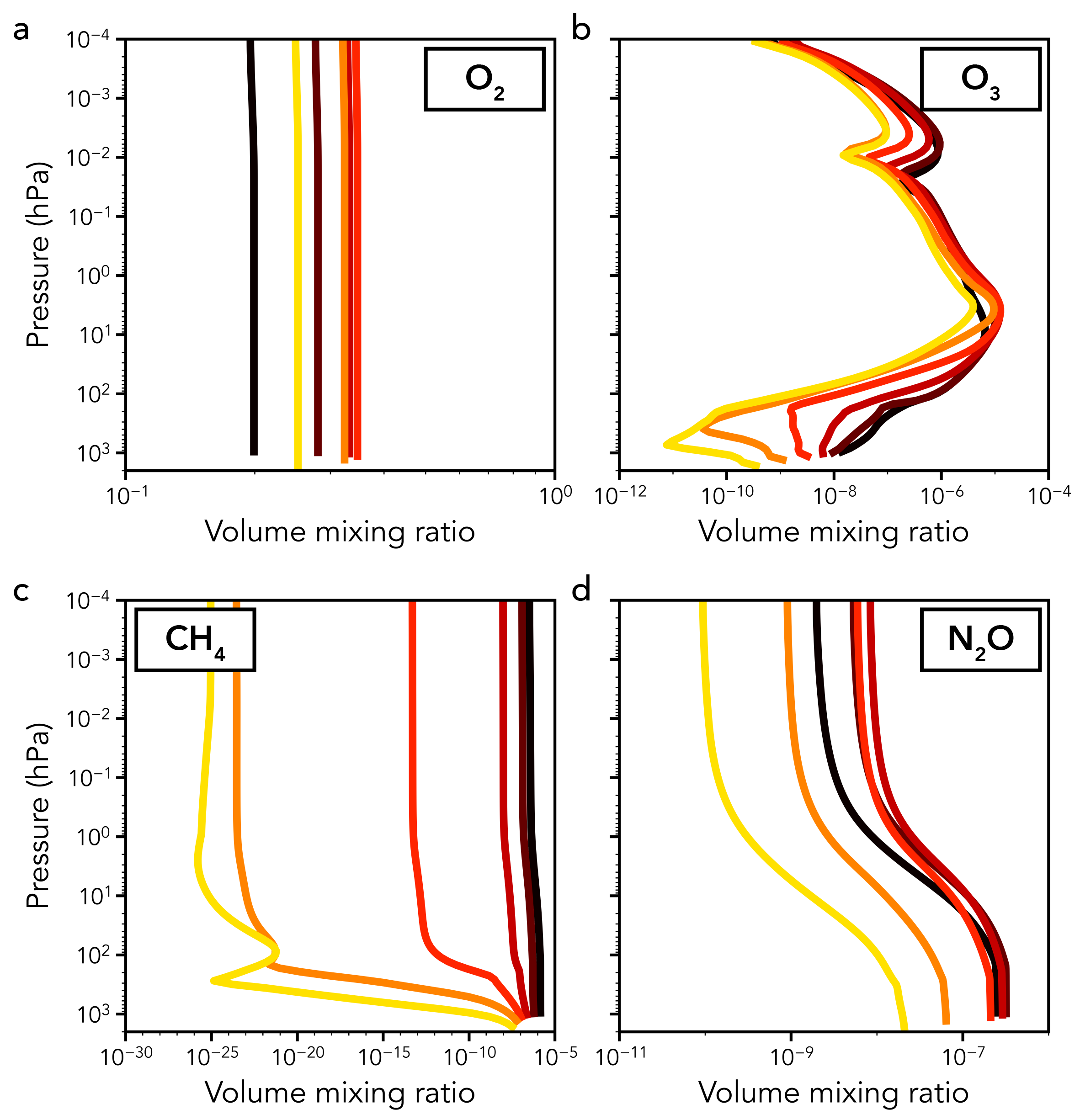}
    \caption{Vertical distribution profiles of the VMRs of O$_2$ (a), O$_3$ (b), CH$_4$ (c),
    and N$_2$O (d) displayed in units of mol/mol for the biotic scenarios (see Fig.~\ref{fig:climate} for the colour legend).}
\label{fig:chem_biosigs}
\end{figure}

In this section, we analyse scenarios 1-6 (with biomass emissions) and describe the climate-chemistry response of atmospheric (potential) biosignatures in our model scenarios. Specifically, we analyse VMR profiles of O$_2$, O$_3$, CH$_4$, and N$_2$O in Figure \ref{fig:chem_biosigs}. 

\subsubsection{O$_2$}
Despite the surface flux of O$_2$ remaining constant (Earth's biomass) for each model run (Table \ref{tab:lower_boundary_flux}), the VMR of O$_2$ steadily 
increases (Figure \ref{fig:chem_biosigs}a) with an increase in solar constant, ranging from 0.20 for $S = 1.0$ to 0.35 for $S = 1.3$. At $S = 1.4$, this
trend reverses, and O$_2$ VMRs fall to 0.32 and then 0.25 mol/mol at $S = 1.5$. From $S = 1.0$ -- 1.5, the
corresponding partial pressures for O$_2$ are 0.20, 0.29, 0.36, 0.42, 0.45, and 0.44 bars respectively - O$_2$ 
production is therefore still increasing in rate up to $S = 1.4$ despite a drop in VMR.
The dominant photochemical sink for
O$_2$ in all model runs is photolysis within the wavelength range $\lambda$ = 10--200 nm above the troposphere. With an increasing solar 
constant, the photolysis rates of O$_2$ steadily increase. However, the presence of more H$_2$O vapour in the simulations with a higher solar
constant introduces a greater abundance of HO$_\mathrm{X}$ species, rising from $<$ 1 pptv for $S = 1.0$ to 1--10 ppbv for $S = 1.4$ 
within regions of 100--10 hPa. This results in catalytic cycles driven by OH, HO$_2$, and H to recycle O($^3$P) atoms (produced from
the photodissociation of CO$_2$, H$_2$O, and O$_3$) into O$_2$ molecules  
increasing in rate faster than the rate of O$_2$ photolysis, enabling the O$_2$ VMR to rise with increasing solar constant. 
At $S = 1.5$, the UV-C radiation flux in the lower atmosphere is orders of magnitude larger due to the falling 
tropospheric O$_3$ 
concentrations, and the O$_2$ photolysis increases in rate faster than O$_2$ is regenerated by the catalytic cycling of 
HO$_\mathrm{X}$ compounds.
Additional O$_2$ is delivered from the top of the 
atmosphere due to the increasing escape rates of
hydrogen and the photolysis of CO$_2$, with the H escape flux reaching 0.69 Tg yr$^{-1}$ for $S = 1.5$. The H escape flux enables the O atoms left behind react with OH and to stimulate the mesospheric
abiotic O$_2$ production without being lost to HO$_\mathrm{X}$
cycles. 

\subsubsection{O$_3$}
\label{subsubsection:ozone}
\begin{figure}[h!]
\centering
    \includegraphics[width=\hsize]{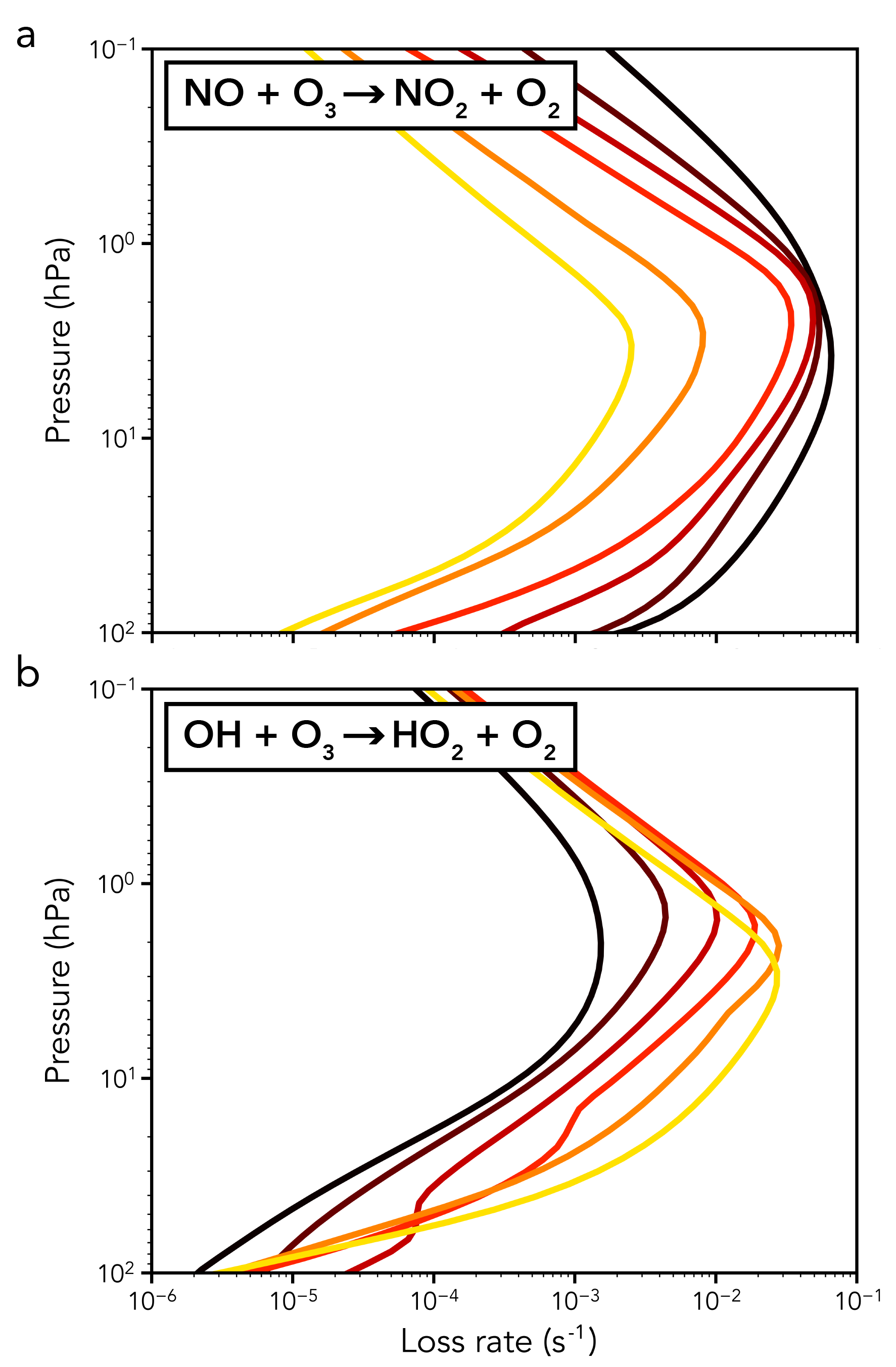}
    \caption{Photochemical loss rate coefficients in the biotic scenarios for the gas-phase removal of O$_3$ with NO (upper panel) and OH (lower panel) for all biotic model scenarios. The reciprocal of the quantity shown gives the atmospheric lifetime in seconds against destruction of O$_3$ by the given reaction (see Fig.~\ref{fig:climate} for the colour legend).}
\label{fig:o3-no-oh-reactions}
\end{figure}

An unexpected result of this work is that the bulk of the atmospheric O$_3$, occurring mainly in the stratospheric ozone layer centred at about 10 hPa (Fig. 2b) mostly survives in all scenarios. 
Why? This result is not due to stronger O$_2$ photolysis stimulating the Chapman cycle, since the necessary UV-C fluxes are not strongly enhanced in the middle stratosphere (Fig. 1e) for the increased instellation scenarios but are instead efficiently absorbed. The survival of stratospheric O$_3$ is all the more puzzling when one notes that UV-B, a photolytic sink for O$_3$, is actually enhanced with increasing instellation (Fig. \ref{fig:climate}d). Results in the stratosphere suggest that HO$_\mathrm{X}$ is strongly increased as expected for the high instellation scenarios with enhanced steam (Figs. \ref{fig:chem_reactive}b and \ref{fig:HOX}). However, the increased HO$_\mathrm{X}$ can act as a sink for NO$_\mathrm{X}$ and ClO$_\mathrm{X}$, and, importantly, this occurs in the regions where these species would otherwise strongly remove ozone catalytically, that is, in the lower stratosphere (NO$_\mathrm{X}$; Figs. \ref{fig:chem_reactive}c and \ref{fig:NOX}) and the mesospheric (ClO$_\mathrm{X}$; Figs. \ref{fig:chem_others}c and \ref{fig:ClOX}). 
Figure~\ref{fig:o3-no-oh-reactions} presents
the loss rate coefficients of the gas-phase reactions involving firstly NO and secondly OH with O$_3$. The reciprocal of the quantity shown indicates the chemical removal lifetime of O$_3$ in seconds via the reaction indicated. This figure suggests that the respective decreases and increases in these two reactions act to maintain an approximately constant
rate of O$_3$ loss with increasing instellation. This response in important O$_3$ sinks is likely the main reason for the survival of stratospheric ozone across the IHZ (despite stronger photolytic UV-B sinks and more catalytic loss from HO$_\mathrm{X}$). A secondary reason for the survival of ozone is the greenhouse-driven stratospheric cooling. This slows the temperature-dependent Chapman sink: O$_3$+O$\rightarrow$2O$_2$. 

The secondary O$_3$ peak in the stratopause at 10$^{-3}$--10$^{-2}$ hPa  remains present for all scenarios, although here more  O$_3$  is removed (with O$_3$ abundances decreased by more than a factor of 10) compared to the primary ozone peak removal in the stratosphere. This result likely arose because at higher altitudes O$_3$ is more efficiently destroyed by catalytic HO$_\mathrm{X}$ cycles \citep{Yung1998-wk}, the 
In the troposphere, O$_3$ decreases with increasing instellation. This occurs despite enhanced OH from the damper atmospheres stimulating the O$_3$ smog cycle production. Further analysis suggested that chlorine (Cl) atoms were important for the O$_3$ removal. The Cl atoms were released from their reservoir HCl, via reaction with OH (steam atmospheres featured high OH abundances) and increased UV-C photolysis of HCl. 

\subsubsection{CH$_4$}
The CH$_4$ abundances decrease with increasing solar constant, as shown in Fig. \ref{fig:chem_biosigs}c. The surface 
concentrations fall from 1.46 ppm to 0.57 ppm, 0.22 ppm, 0.11 ppm, 0.06 ppm, and 0.03 ppm for $S = 1.0, 1.1, 1.2, 1.3, 1.4,$
and $1.5,$ respectively. Larger H$_2$O abundances lead to an increase in OH by more than two orders of magnitude in the lower and middle atmosphere (Fig. \ref{fig:HOX}), 
which combines with the increasing rate of CH$_4$ photolysis in the upper layers to increase the overall rate of CH$_4$ oxidation in the atmosphere,
effectively increasing the abundances of methane's main oxidation products  CO$_2$ and H$_2$O throughout the atmospheric column. Chlorine atom (Cl)
abundances increase to maximum values of 5.43, 42.65, 68.11 pptv in the troposphere for $S = 1.3, 
1.4,$ and $1.5,$ respectively, compared to the 0.001--0.1 pptv abundances seen for $S = 1.0$--$1.2$ (Fig. \ref{fig:ClOX}) as already discussed. The 
additional Cl atoms greatly increase the rate of CH$_4$'s chemical breakdown closer to the surface, resulting in the 
substantially larger decreases in the CH$_4$ column for $S = 1.3$--$1.5$ (Fig. \ref{fig:chem_biosigs}c) scenarios.

\subsubsection{N$_2$O}
The N$_2$O abundances rather surprisingly increase in the upper (P $<$ 10 hPa) atmosphere as the solar constant increases from the S=1.0 to 1.3 scenarios (Fig. 
\ref{fig:chem_biosigs}). This behaviour is unexpected because increasing instellation is an important photolytic sink for N$_2$O, as is the case for $S = 1.4$ and $1.5$, whereby the N$_2$O abundances over all altitudes decrease to values below those at $S = 1.0$. The chemical budget of N$_2$O in our model is dominated by (1) biological emission at the surface (held constant in all scenarios), (2) eddy mixing throughout the atmospheric column, which controls the rate of transport of (biomass) N$_2$O from the lower levels up to the middle atmosphere, where it is destroyed in situ, (3) gas-phase removal via photolysis in the UV-B or/and reaction with electronically excited oxygen atoms mainly in the middle atmosphere and above, and (4) minor (<1 percent) in situ abiotic, gas-phase sources. Further analysis suggested that the N$_2$O mesosphere increase for the S=1.0 to 1.3 scenarios was related to a modest decrease in electronically excited singlet oxygen atoms (O($^1$D)), which are a sink for N$_2$O. This, in turn arose due to less O$_3$ (a photolytic source for O($^1$D)) related to HO$_\mathrm{X}$ destruction in the enhanced steam atmosphere at pressures of 10$^{-3}$--10$^{-2}$ hPa, and more H$_2$O, which reacts with O($^1$D) to produce two OH molecules. Cl atoms also act as a chemical sink for N$_2$O. For S $<$ 1.3, the Cl atom abundances in the troposphere stay below 0.1 pptv
and have negligible impact on this biosignature. With the increasing rate of HCl photolysis for S $\geq$ 1.3, the Cl abundances range from
1--100 pptv and take over as the primary atmospheric sink for N$_2$O in the 
troposphere. This additional chemical sink explains
the change in tendency for N$_2$O as the VMRs begin to fall below those calculated at $S = 1.0$ in 
the lower atmosphere, but initially increase when pressures are below 1 hPa.

\subsection{Response of CO$_2$, hydrogen, and nitrogen reservoirs}

\begin{figure}[h!]
\centering
    \includegraphics[width=\hsize]{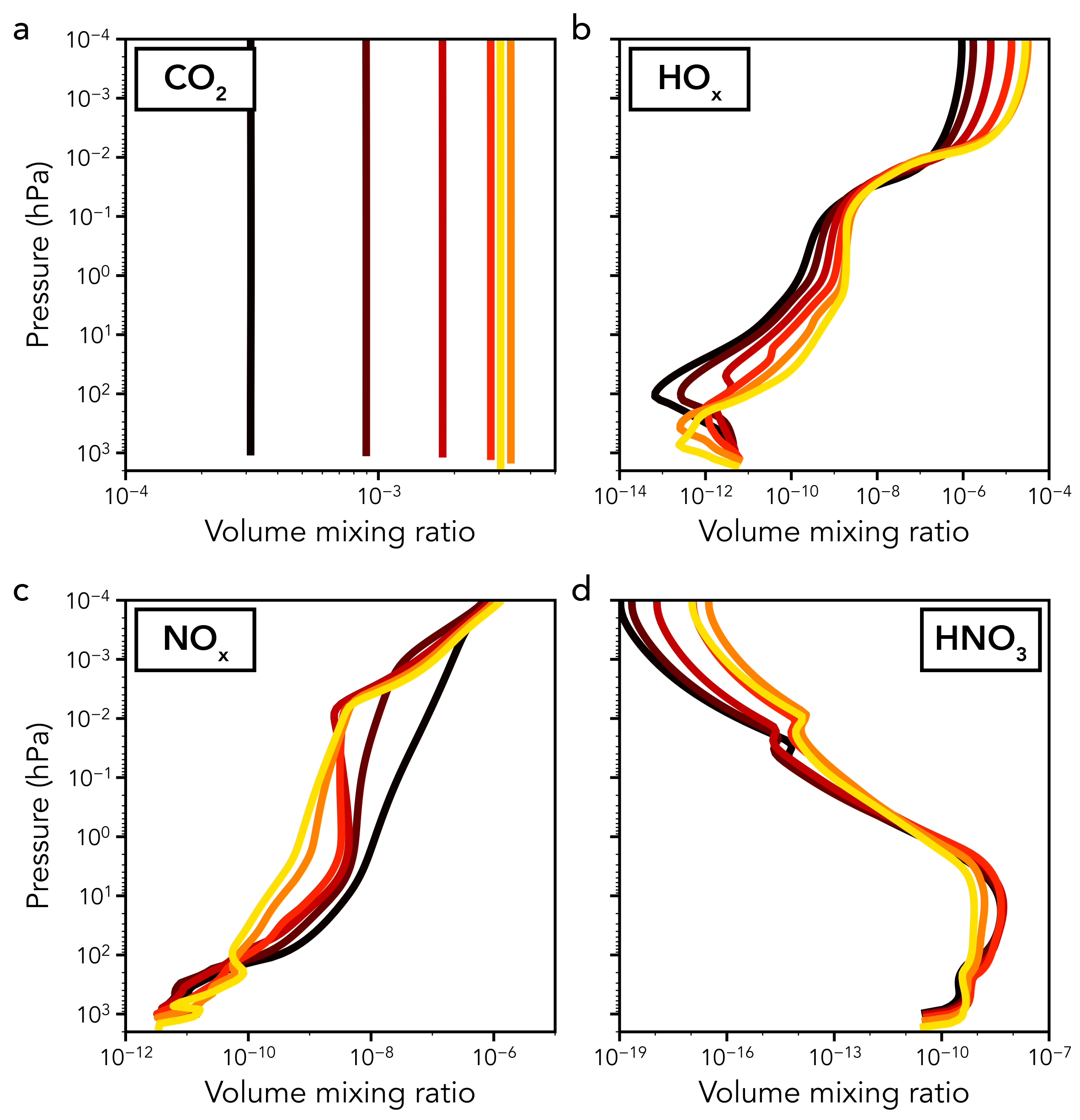}
    \caption{Vertical distribution profiles of the VMRs of CO$_2$ (a), the HO$_\mathrm{X}$ group (b; see also Fig.~\ref{fig:HOX}), the NO$_\mathrm{X}$ group (c; see also Fig.~\ref{fig:NOX}),
    and HNO$_3$ (d) displayed in units of mol/mol for the biotic scenarios (see Fig.~\ref{fig:climate} for the colour legend).}
\label{fig:chem_reactive}
\end{figure}

\subsubsection{CO$_2$ and HO$_\mathrm{X}$}
Figure \ref{fig:chem_reactive} presents the VMR profiles of carbon dioxide (CO$_2$), the 
odd-hydrogen group (HO$_\mathrm{X}$ = H + OH + HO$_2$; see also Fig. \ref{fig:HOX}), the odd-nitrogen group (NO$_\mathrm{X}$ = N + NO +
NO$_2$ + NO$_3$; see also Fig. \ref{fig:NOX}), and nitric acid (HNO$_3$), an important reservoir species for HO$_\mathrm{X}$ and NO$_\mathrm{X}$. With an increasing solar constant and a constant
surface flux (Table \ref{tab:lower_boundary_flux}), CO$_2$ VMRs increase from 310.0 ppm at $S = 1.0$ to
3368.4 ppm at $S = 1.4$ while maintaining an isoprofile vertical structure (suggesting that chemical timescales are longer than those of transport), but begin to fall at $S = 1.5$
to a value of 3064.0 ppm. With increasing proximity
to the Sun, the UV photolysis of CO$_2$ to CO and O($^3$P) or O($^1$D) atoms rises throughout the atmosphere.
This is accompanied by an increasing abundance of HO$_\mathrm{X}$ species produced from the increasing
H$_2$O abundances seen in Fig. \ref{fig:climate}b being produced by the rising rate of photolysis for
H$_2$O. The atmospheric production of CO$_2$ is dominated by the catalytic recycling of CO by the hydroxyl
radical \citep[OH;][]{Yung1998-wk}, which becomes more powerful with the growing HO$_\mathrm{X}$ abundances as solar constant 
values rise. This increasing production strength exceeds that of the increasing photolysis of CO$_2$,
resulting in CO$_2$ increasing in abundance with increasing S. For $S = 1.5$ however, this effect reverses, since the
increased photolysis rate of CO$_2$ now exceeds the rising rate of CO-OH recombination.

Figure \ref{fig:HOX} displays the VMRs of the individual species family members H, OH, and HO$_2$ to the HO$_\mathrm{X}$
vertical distribution profile. The total HO$_\mathrm{X}$ concentrations, where pressures exceed 100 hPa 
(Earth's tropopause), increases with rising solar constant up to but not including $S = 1.3$. These increases
are due to the additional OH being introduced by both enhanced H$_2$O VMRs as well as increasing 
UV fluxes. For S $\geq$ 1.3, however, the rising UV flux notably increases the abundance of Cl atoms close
to the surface through the photolytic breakdown of HCl (see Fig. \ref{fig:ClOX}a). The higher Cl abundances lead to the removal of low-altitude HO$_2$ that replenishes 
HCl and O$_2$ faster 
than they can be produced by the breakdown of the additional atmospheric H$_2$O. The increasing UV-C flux (Fig.
\ref{fig:climate}e), a product of the strong decrease in tropospheric O$_3$ (Fig. \ref{fig:chem_biosigs}b, described in greater detail in Sect. 3.3.2), also decreases the stability of HO$_2$ near the
surface due to an increase in its rate of photolysis. HO$_2$ is the dominant HO$_\mathrm{X}$ species
at pressures greater than 10$^2$ hPa and the HO$_\mathrm{X}$ profiles of Fig. \ref{fig:chem_reactive} therefore show a decrease close to
the surface with increasing S. For atmospheric pressures between 10$^{-1}$ and 10$^2$ hPa, OH becomes the primary species in the HO$_\mathrm{X}$ 
family. As it is the direct product of H$_2$O reactions with
O($^1$D) atoms and photolysis the model runs feature enhancements in mid-altitude OH 
abundances as  expected, increasing from VMRs of approximately 0.1 pptv at $S = 1.0$ to 100 pptv at $S = 1.5$. This explains the steady rise in the 
overal HO$_\mathrm{X}$ group in the middle atmosphere. 

At high altitudes (pressures $<$ 0.1 hPa), H atoms are the primary
constituent of HO$_\mathrm{X}$, related to the high rates of UV induced photolysis at the top of the model atmosphere. Between
0.01 and 1 hPa in Fig. \ref{fig:HOX}, however, the H atom abundances do not vary substantially with increasing solar constant. 
The O$_2$ VMRs vary between 0.20 and 0.35 mol/mol (Fig. \ref{fig:chem_biosigs}a) and provide H atoms with a roughly steady chemical
sink as they react to produce HO$_2$. The stability of this chemical sink balances the increasing production of H atoms from
the elevating H$_2$O abundances between 0.01 and 1 hPa, resulting in the approximately invariant HO$_\mathrm{X}$ profile
here across $S = 1.0-1.5$. At higher altitudes up to the model lid (10$^{-4}$ hPa), the photolysis of H$_2$O and subsequent
production of H atoms grows larger than the O$_2$ sink can balance, producing an enhanced H atom abundance with increasing solar constant.

\subsubsection{NO$_\mathrm{X}$ and its reservoirs}
The NO$_\mathrm{X}$ species (Fig. \ref{fig:chem_reactive}c) exhibit three different trends in the lower (P $\geq$ 100 hPa), 
middle (100 $>$ P $\geq$ 0.01 hPa), and upper (P $<$ 0.01 hPa) atmosphere. In the lower atmosphere, NO$_\mathrm{X}$ abundances 
increase with increasing solar constant. A fixed surface flux of N$_2$O, coupled with an increasing surface flux of UV radiation, increases the 
rate of NO production via N$_2$O photolysis in the troposphere. Additionally, a decreasing HO$_2$ abundance close to the surface at high S lowers the chemical sink 
strength for NO, which is the most abundant NO$_\mathrm{X}$ family member in the lower atmosphere, and its increasing production and
decreasing sink result in an increase in the NO$_\mathrm{X}$ VMR profile. In the middle atmosphere, NO$_\mathrm{X}$ abundances
fall with an elevating solar constant by between 1 and 2 orders of magnitude. This is related to the conversely
increasing HO$_\mathrm{X}$ (Fig. \ref{fig:chem_reactive}b). 

In
the mesosphere, the rising photolysis rate of N$_2$ with increasing solar constant leads to an enhancement in the abundance of N atoms, which quickly convert to NO$_\mathrm{X}$ species. The resulting NO$_2$ molecules are
 converted to HNO$_3$ by the enhanced HO$_\mathrm{X}$ abundances, resulting in a decrease in NO$_\mathrm{X}$
for an increasing S value until the photolytic frequencies of $S = 1.4$ and $1.5$ are large enough to make N atoms the dominant
NO$_\mathrm{X}$ compound. When this occurs, the  NO$_\mathrm{X}$ concentration becomes roughly invariant above 
altitudes where pressures are less than 10$^{-3}$ hPa.

\subsection{Response of O($^1$D), chlorine chemistry, and CO}
\begin{figure}
\centering
    \includegraphics[width=\hsize]{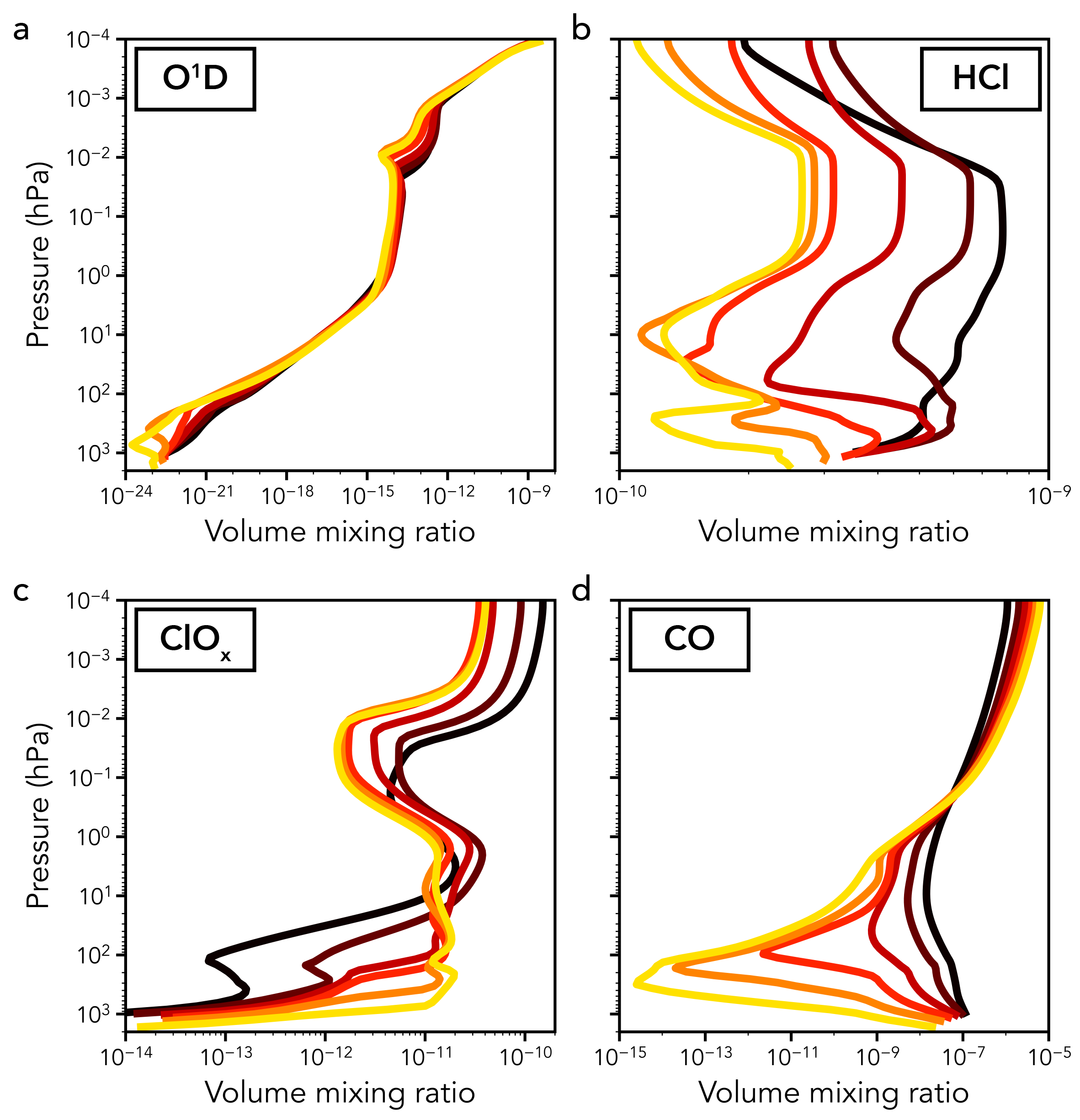}
    \caption{Vertical distribution profiles of the VMRs of O($^1$D) (a), HCl (b), the ClO$_\mathrm{X}$ group (c; see also Fig. \ref{fig:ClOX}),
    and CO (d) displayed in units of mol/mol for the biotic scenarios (see Fig.~\ref{fig:climate} for the colour legend).}
\label{fig:chem_others}
\end{figure}

The VMR profile of O($^1$D) is presented in Fig.
\ref{fig:chem_others}a. At pressures greater than 100 hPa, the concentrations decrease
by two orders of magnitude with increasing solar constant. Between 1 and 100
hPa, the concentrations remain approximately stable as the solar constant rises. In the region 
between $\approx$0.01 and 0.1 hPa the O($^1$D) abundances again experience 
drops from maximum VMRs of 0.27 pptv at $S = 1.0$ to 
3.44$\times$10$^{-3}$ pptv at $S = 1.5$. At the top of the atmosphere, where 
pressures are below 10$^{-3}$ hPa, the concentrations remain stable. The principle
source of O($^1$D) is the UV photolysis of O$_3$ within the UV-C wavelength
range. CO$_2$ photolysis also provides a source of O($^1$D), but is less
efficient due to its smaller cross-sectional area within the UV-C band. The 
most efficient sink for O($^1$D) is collision induced de-excitation with N$_2$,
converting O($^1$D) to O($^3$P) . N$_2$ remains
the primary component of the atmosphere (0.78 mol/mol at $S = 1.0$ and falling
to 0.41 mol/mol at $S = 1.5$). In general, the behaviour of O($^1$D) follows the behaviour
of O$_3$, detailed in Sect. \ref{subsubsection:ozone}.

The source of chlorine in all model runs is the surface flux (62.38 Tg yr$^{-1}$) 
and volcanic fluxes  of HCl (4.30 Tg yr$^{-1}$) and the surface flux 
of CH$_3$Cl (2.95 Tg yr$^{-1}$). The primary sinks of HCl are reactions with OH,
and UV photolysis with cross-section maxima centred on 150--160 nm within the 
UV-C radiation range (Fig. \ref{fig:climate}e). At pressures greater than 100 hPa,
HCl concentrations remain quite stable for instellations of $S = 1.0-1.2$. UV-C fluxes fall 
between $S = 1.0$ and $1.2$, whereas OH gradually increases. This results in HCl maintaining 
concentrations of 0.50--0.60 ppbv for runs at $S = 1.0, 1.1$, and $1.2$. Through $S = 
1.3-1.5$, the 
emerging UV-C window near to the surface and the still increasing OH concentration,
increase the rate of HCl destruction and its concentration decreases from 0.35 ppbv to
0.12 ppbv. At pressures less than 100 hPa, HCl tends to decrease in abundance 
gradually with increasing S. Increasing OH concentrations and rising UV fluxes 
increase the rate of HCl destruction, but the rise in HO$_2$ enables HCl to
be reproduced through reactions with Cl atoms, which mitigates the decrease in
atmospheric HCl.
 
Figure \ref{fig:chem_others}c presents the ClO$_\mathrm{X}$ (= Cl + ClO + 
ClO$_2$) profiles of the various scenarios (see also Fig. \ref{fig:ClOX}). The primary source of ClO$_\mathrm{X}$ 
is the photolysis of HCl, to produce Cl atoms, and reactions of HCl with OH to 
produce ClO. ClO$_2$ is then produced by the three-body reaction of Cl with O$_2$.
Figure \ref{fig:ClOX} presents the individual profiles of the ClO$_\mathrm{X}$
group members. At pressures greater than 100 hPa OH dominates HCl loss between
S = 1.0 and 1.3, resulting in ClO
being the dominant member of the ClO$_\mathrm{X}$ group with concentrations
of 0.1--100 pptv (Cl = 10$^{-4}$--10$^0$ pptv). When the UV-C window
appears at $S = 1.4-1.5$, photolysis dominates HCl loss and atomic Cl becomes the 
main component of ClO$_\mathrm{X}$ with abundances of 1--10 pptv.

\subsection{Comparisons to planet scenarios with no active biosphere}

\begin{figure}
\centering
\includegraphics[width=\hsize]{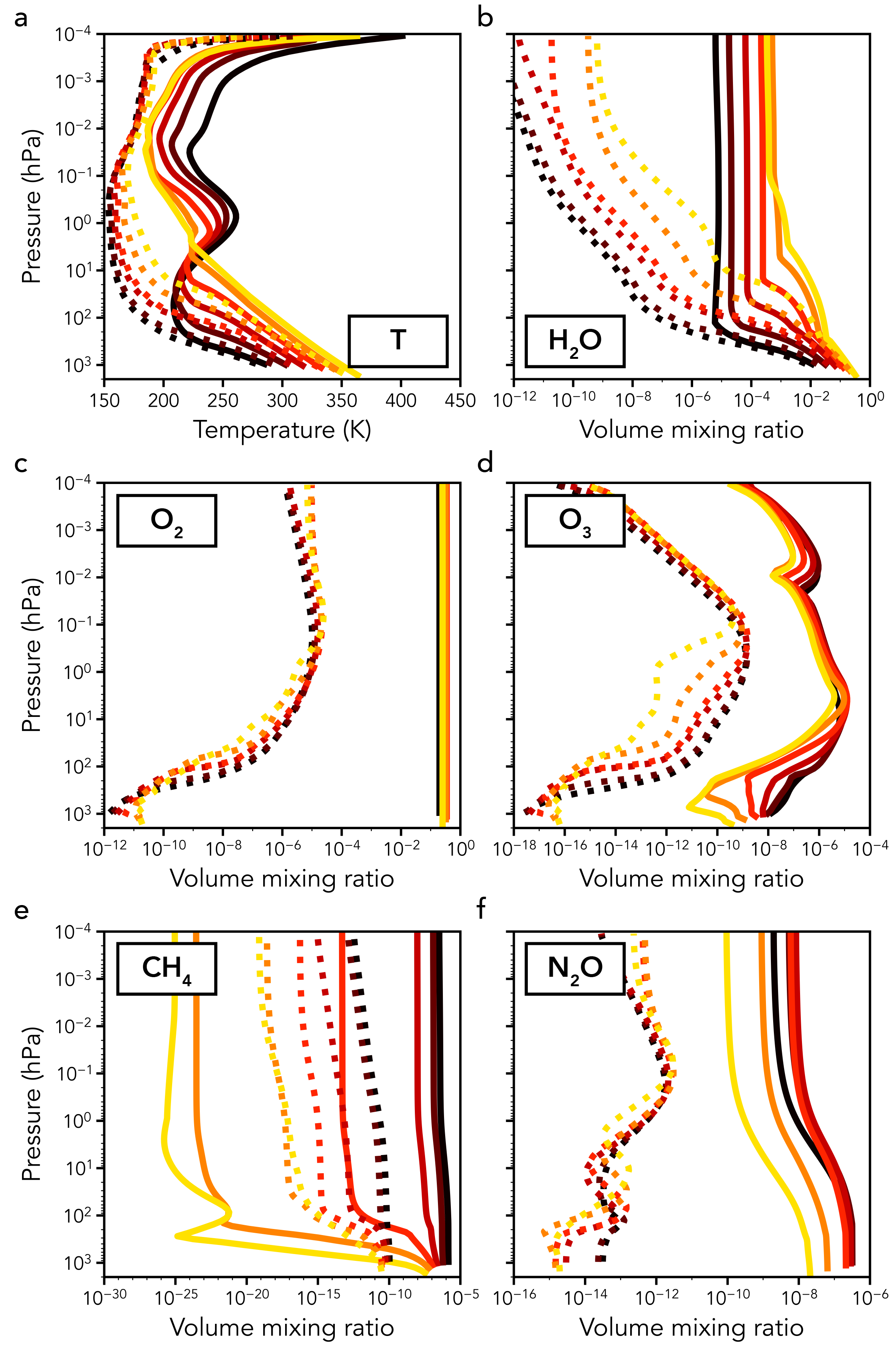}
\caption{Profiles produced by 1D-TERRA for planets with (solid lines) and without (dashed lines) Earth's biomass emissions at the surface. The dashed lines represent the abiotic runs. Line colours have the same meanings as in Figs. \ref{fig:climate}--\ref{fig:chem_others} and denote the increasing
solar instellation.
}\label{fig:BiovsAbio_Profiles}
\end{figure}

In addition to the six biotic scenarios shown in Table \ref{tab:lower_boundary_flux}, we
performed simulations across the same range of instellations with only volcanic outgassing
fluxes as input. These scenarios represent an Earth-like planet without any biological
emissions at the surface. The surface relative humidity for all scenarios was again fixed to 
80\%, with layers below the cold trap set to the Manabe-Wetherald profile \cite{manabe1967thermal} and layers above calculated via photochemistry and diffusion. This
is to simulate the presence of an ocean. For simplicity, the negative surface flux of CO$_2$ 
in Table \ref{tab:lower_boundary_flux} is tuned to produce a surface CO$_2$ abundance 
approximately equal to the $S = 1.0$ biotic value ($\sim$ 310 ppm).
A lower negative surface flux would be expected due to an abiotic surface having no land vegetation for
photosynthesis and a lack of a biologically driven carbonate-silicate cycle
in the oceans, but true consideration of these processes is beyond the scope of the present study. This
important caveat is expanded upon in Sect. \ref{sec:discussion}.

Figure \ref{fig:BiovsAbio_Profiles} presents vertical profiles of the atmospheric temperature
and important atmospheric species of the abiotic planet simulations performed by the
1D-TERRA climate-chemistry model compared to the biotic runs. In Fig. \ref{fig:BiovsAbio_Profiles}a, the 
abiotic temperature profiles (dashed lines) from $S = 1.0$--$1.3$ do not produce the 
temperature inversion between 0.1 and 1 hPa seen in the biotic simulations
(solid lines) due to the weakened ozone layer. Surface temperatures reach 280.90 K, 293.86 K,
308.50 K, 332.51 K, 346.94 K, and 364.38 K for $S = 1.0$--$1.5$.
For $S \geq 1.4$, the classical greenhouse effect produced by the rising
tropospheric water content (Fig. \ref{fig:BiovsAbio_Profiles}b) results in the
cold trap rising in altitude as atmospheric temperatures increase. At this point,
the water vapour within the troposphere diffuses to the higher altitudes and
becomes more comparable in both vertical and column abundance to the biotic simulations
below 10 hPa. The cooling of the stratosphere that occurs in the biotic scenarios
is not seen in the abiotic scenarios, producing strong temperature differences between biotic
and abiotic runs across $S = 1.0--1.3$ at altitudes with pressures greater than 0.1 hPa. 

The lower stratospheric H$_2$O in the abiotic scenarios that arises due to a lower CH$_4$ surface flux
is less efficient at trapping the heat travelling upwards, enabling the stratosphere to gradually heat. Without an efficient O$_3$ layer (Fig. \ref{fig:BiovsAbio_Profiles}d) absorbing UV radiation,
photolysis rates of H$_2$O in the stratosphere and mesosphere are greater in the
abiotic scenarios. At altitudes where pressures range from 10$^{-2}$--10 hPa, the biotic scenario H$_2$O
chemical lifetime ranges between 1 and 10 years. In the abiotic scenarios, these values fall to 0.01--0.1 years
due to the greater photolysis rates. The combination of a lower photochemical source in CH$_4$, an increased photochemical sink, and the continued suppression of H$_2$O diffusion from within the troposphere due to the cold 
trap cause our abiotic scenarios to produce very low mesospheric H$_2$O abundances.

Without the biotic emission of O$_2$ at the surface (Table \ref{tab:lower_boundary_flux}),
the abiotic runs produce O$_2$ via atmospheric chemistry only. CO$_2$ photolyses into oxygen
atoms, which are capable of combining into molecular O$_2$. As the instellation increases, we see
the surface VMRs rise from 1--10 pptv (Fig. \ref{fig:BiovsAbio_Profiles}c), significantly lower than the biotic runs. As O$_3$ is dependent on the O$_2$ abundances, it is produced
at levels up to 6 orders of magnitude lower than the biotic counterparts (\ref{fig:BiovsAbio_Profiles}d). The O$_3$ layer (abundances between 0.01 and 1 ppbv) in the abiotic runs is maintained
by the same balance between NO$_\mathrm{X}$ and HO$_\mathrm{X}$ groups seen in
the biotic runs (Fig. \ref{fig:o3-no-oh-reactions}) up to and including 
instellations of $S = 1.3$. At $S = 1.4$, the HO$_\mathrm{X}$ concentrations become
three orders of magnitude greater in this pressure region in the abiotic runs
than the biotic runs. This increases the sink rate of O$_3$, resulting in
the abiotic O$_3$ layer collapsing from ppbv abundances to 10$^{-2}$--10$^{-1}$ 
pptv abundances. 

There are three to four orders of magnitude less methane (Fig. \ref{fig:BiovsAbio_Profiles}e) than their biotic counterparts in the abiotic simulations between instellations $S = 1.0$ and $1.3$ due to the removal of the biomass emissions. The 
lower CH$_4$ abundances result in a lower photochemical production of H$_2$O above the
troposphere. By $S = 1.4$ vertical diffusion from the troposphere dominates as the source
of stratospheric and mesospheric H$_2$O resulting in their differences between the biotic scenario
decreasing. As a result, the temperature profiles
of the biotic and abiotic simulations become comparable below the troposphere
for runs where $S \geq 1.4$ (Fig. \ref{fig:BiovsAbio_Profiles}a).

\subsection{Emission spectra responses of biosignatures to increasing stellar instellation}

\begin{figure}
\centering
\includegraphics[width=\linewidth]{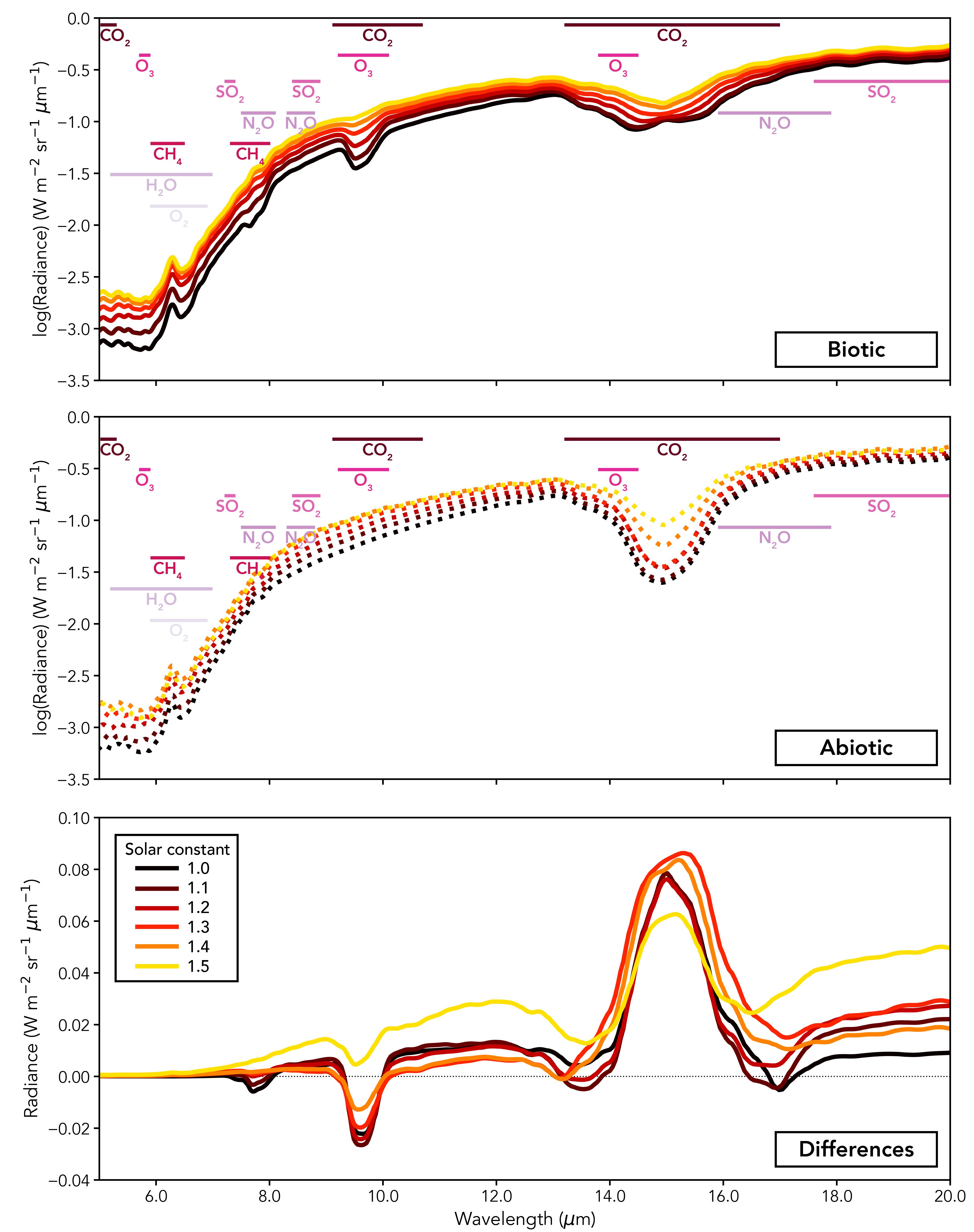}
\caption{Emission spectra of the six planetary instellation scenarios produced by the
GARLIC radiative transfer model. The spectra are produced using a spectral
resolving power of R = 100 across a wavelength range of 5.0--20.0 $\mu$m. (a) Scenarios with Earth biomass emissions at the surface. (b) Scenarios without Earth biomass emissions at the surface. (c) Difference between the
biotic and abiotic scenario fluxes at each respective instellation.}\label{fig:emission}
\end{figure}

Figure \ref{fig:emission} presents the resulting emission profiles of the six planet 
scenarios driven with and without Earth's biomass surface fluxes. This model takes as input converged 
temperature, pressure and composition profiles from the 1D-TERRA model and molecular absorption features
from the HITRAN line-by-line database \citep{hitran2020}. The spectral resolving power is set to R = 100.

From 5.0--7.1 $\mu$m the H$_2$O absorption band dominates the emission spectra in both
biotic and abiotic scenarios. With increasing instellation, the increasing H$_2$O abundances
in the lower atmosphere dominate the spectral features of the biotic scenarios and their
abiotic counterparts. The significantly stronger CH4$_4$ and O$_2$ abundances in the biotic 
runs (Figs. \ref{fig:BiovsAbio_Profiles}c and d) produce marginally stronger absorption
features at 6.0--6.5 $\mu$m. The first noticeable difference between the biotic and abiotic 
emission spectra is seen at 7.8--8.0 $\mu$m for cases $S = 1.0$--$1.2$. These are produced by the presence
of larger CH$_4$ and N$_2$O abundances in the biotic cases - as seen in Figs. \ref{fig:BiovsAbio_Profiles}c and f,
the abundances in the biotic scenarios fall to negligible concentrations for $S > 1.2$.

Between 9.0 and 10.0 $\mu$m, the O$_3$ layer produced by the biotic surface fluxes of O$_2$ are distinguishable
throughout all cases of $S$. As the O$_3$ abundances in abiotic scenarios are typically 4--6 orders of 
magnitude lower than their biotic counterparts, the feature on this wavelength range remains as a strong
distinguishing signature of biological surface activity. The O$_3$ signature located at 9.6 $\mu$m in Fig. 
\ref{fig:emission} remains apparent for runs with instellations up to
and including a value of 1.4 - a distance corresponding to 0.85 AU to the Sun. At $S = 1.5$, the O$_3$ profiles begin 
to recede in the middle atmosphere (Fig. \ref{fig:chem_biosigs}b) and the tropospheric content has fallen to 
magnitudes
under 1 ppbv, which results in the CO$_2$ absorption feature in this wavelength range smearing out that
of O$_3$. A similar effect is seen for the O$_3$ feature found at 13.8--14.8 $\mu$m. It remains apparent in the
GARLIC spectra up to and including runs at $S = 1.5$, but becomes weakened by the competing CO$_2$ absorption band
across this wavelength interval.

From 10-0--13.0 $\mu$m, the differences in
the atmospheric temperature profiles of biotic and abiotic runs (Fig. \ref{fig:BiovsAbio_Profiles}a) begin to 
make an impact on the spectra. Approximately 99\% of the atmospheric column is contained at pressures above
10 hPa, causing this altitude region to dominate the contributions to the emission spectra. 
The greater H$_2$O abundances in the stratosphere in biotic cases across $S \leq 1.3$, an oxidation product of the 
biotic CH$_4$ surface fluxes that the abiotic cases lack, produces biotic-abiotic temperature profile differences at 
pressures $>$ 10$^1$ hPa that result in greater emission fluxes across 10--13.0 $\mu$m. At the collapse of
the cold trap ($S \geq 1.4$), the H$_2$O abundances below 100 hPa become approximately equal in the biotic and 
abiotic scenarios, resulting in temperature profiles that are harder to distinguish, and the differences in
their emission features at 10--13.0 $\mu$m begin to fall.

\subsection{Detectability of biomarkers with LIFEsim}

\begin{figure*}
\centering
\includegraphics[width=\textwidth]{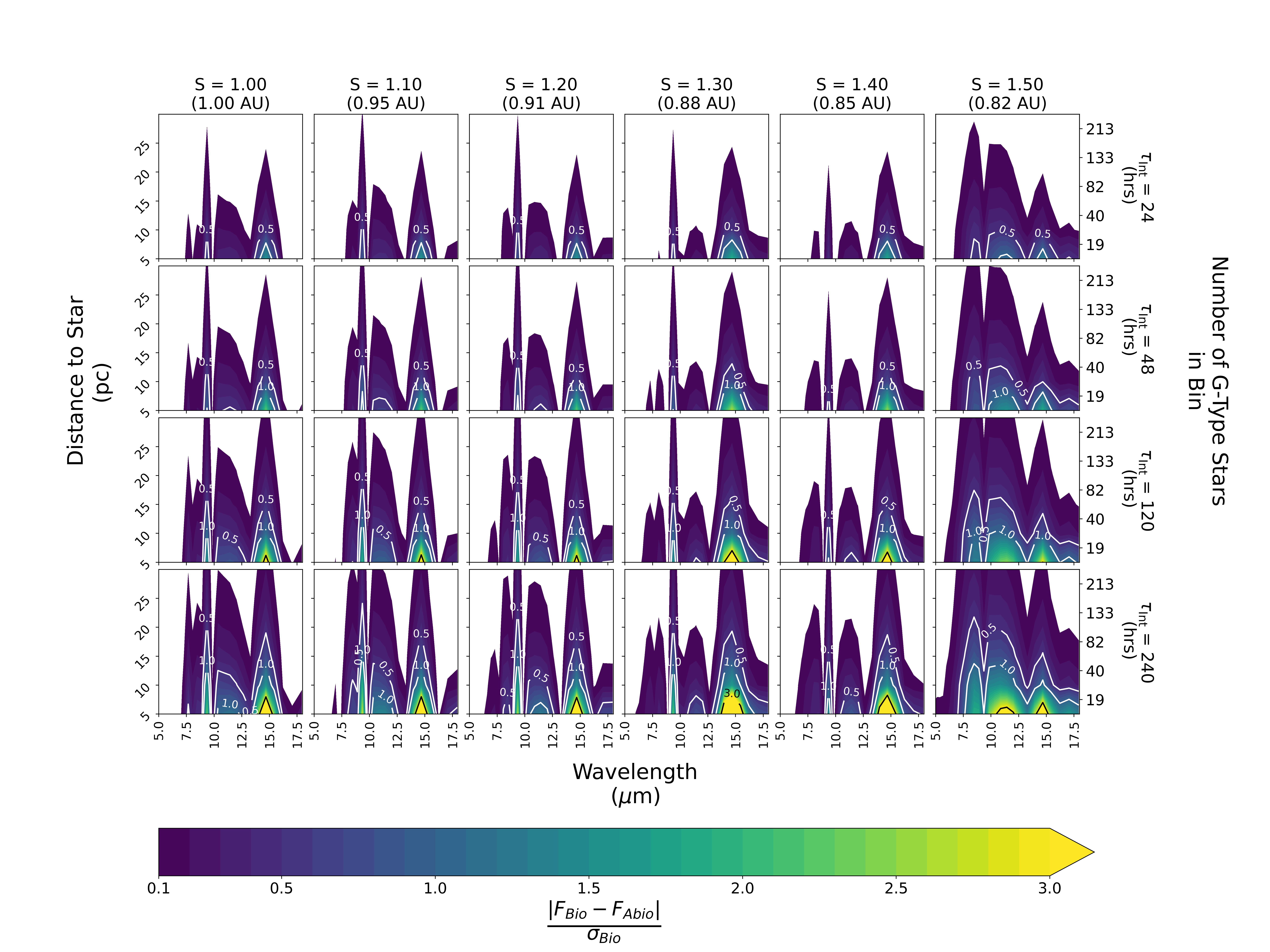}
\caption{Statistical significance (as a function of wavelength) of the difference in retrieved emission flux calculated via LIFEsim between the biotic and abiotic scenarios studied in this work. The columns (from left to right) show 
increasing planet instellation; the different rows show different observation integration times (from 24--240 hours)}\label{fig:LIFEsim_BiovsAbio}
\end{figure*}

LIFEsim, developed by \cite{dannert2022lifesim} and used across works such as \cite{quanz2022large}, \cite{konrad2022large}, and \cite{alei2022earthanalog}, was used here to
add astrophysical noise signals to the emission
spectra produced by the GARLIC radiative transfer code. The absolute difference
in received flux from the biotic and abiotic scenarios at each instellation is found
at each wavelength interval, and divided by the biotic spectra's noise level.
These calculations are performed over a stellar distance grid of 5--30 pc in 
intervals of 5 pc, and for LIFE observation times of 24, 48, 120, and 
240 hours using the `baseline' scenario. This uses a mirror diameter of 2.0 m, system throughput
of 5\%, quantum efficiency of 70\%, and 
a spectral resolving power of R = 20. Figure \ref{fig:LIFEsim_BiovsAbio} presents
the resulting statistical significance of the emission features of this work's
biotic scenarios compared to their corresponding abiotic scenarios. The angular separation
of each planet with varying stellar instellation ($a$, converted from AU to pc) at each 
system distance ($d$, in pc) is calculated via $\tan^{1}$($a/d$) and used as LIFEsim input.

Water features in the IR (5.0--7.1 $\mu$m; Fig. 
\ref{fig:emission}) are indistinguishable between the biotic and abiotic scenarios. 
Likewise, the CH$_4$ band
at 7.7$ \mu$m from the biotic runs is not strong enough to produce a feature with
statistical significance greater than 0.1 with respect to the abiotic planet spectra
for any integration time considered. Increasing the aperture diameter of LIFE to 3.5 m (LIFEsim's `optimistic' setting; Fig. \ref{fig:LIFEsim_BiovsAbio_Optimistic}) and increasing the system throughput and
quantum efficiency to a more optimistic 20\% and 80\%, respectively, enables the
statistical significance of the 7.7$\mu$m CH$_4$ feature to exceed 0.5 for systems within 7.5 pc
for integration times above 120 hours, and $\sim$ 12.5
pc for 240 hours.

From Fig. \ref{fig:LIFEsim_BiovsAbio}, the most distinguishable features
between biotic and abiotic scenarios are the O$_3$ band found at 9.6 $\mu$m and the CO$_2$ (and thermal
emission) band at $\sim$15.0 $\mu$m. For O$_3$ at 9.6 $\mu$m, biotic O$_3$ signals from 
Earth-size planets within the
habitable zone (0.95--1.37 AU, as parameterised by \citealt{kasting1988runaway}) only become
statistically differentiable (liberally defined here as $\geq$ 1.0) from abiotic Earth-like planets
for integration time exceeding 120 hours for systems within 12.5 pc when LIFEsim's baseline scenario 
was used. The optimistic aperture diameter enables
statistical significance of the 9.6 $\mu$m to exceed 1.0 for systems 15--20 pc from Earth
for integration times above 120 hours (Fig. \ref{fig:LIFEsim_BiovsAbio_Optimistic}).

Across 14.0--16.0 $\mu$m, the thermal emission band overlaps the CO$_2$ absorption band. As
the biotic and abiotic runs hold approximately equal CO$_2$ concentrations, the 
differences in this region are produced by the differing temperature profiles (Fig. 
\ref{fig:BiovsAbio_Profiles}a) that also induce temperature broadening of the CO$_2$ bands in the biotic. The ability to distinguish biotic and abiotic runs 
with instellations S $>$ 1.4 is hindered due to the collapse in the cold trap and the
rush of H$_2$O into the stratosphere in both the biotic and abiotic scenarios (Fig. 
\ref{fig:BiovsAbio_Profiles}b). With LIFEsim's baseline scenario, the thermal profile
differences produce emission spectra statistical significances greater than 0.5 for planets
within 12.5 pc after 48 hours of observation. Increasing the integration time to 120 and 240
hours raises this distances to 15--20 pc for planets with $S =$ 1.0--1.5. Within
the habitable zone ($S \leq$ 1.1), values of 1.0 can be reached for stars within 12.5 pc after 120 hours.


\section{Discussion}\label{sec:discussion}

This modelling study does not account for changes to CO$_2$ induced by an active carbonate-silicate cycle,
a vital regulator of atmospheric temperature for Earth's climate. With increasing surface temperatures,
the weathering rate of silicate rocks will increase \citep{west2005tectonic} and draws more gaseous CO$_2$ from
the atmosphere effectively increasing the CO$_2$ surface sink with increasing instellation. Despite this omission,
our biotic scenario with 10\% more instellation produces a surface temperature of 304.41 K. This corroborates
the prior findings of \cite{caldeira1992life} who used a climate model with biologically mediated silicate weathering to study Earth's biosphere as the Sun's luminosity rises who modelled a surface temperature of 305 K 
at $S$ = 1.1. \citet{ozaki2021future} developed a combined biogeochemistry and climate model of Earth's biosphere to 
forward-model the atmospheric responses through time. They found that in approximately 1 Gyr, when S $\simeq$ 1.1, 
atmospheric O$_2$ may drop below 1\% of the present atmospheric level. With the expected decreasing CO$_2$ 
abundances, the productivity of O$_2$ generating biotic activity becomes limited, which would suppress the current 
surface flux of O$_2$ with time and increasing solar constant. The surface fluxes of O$_2$ and CO$_2$ are 
therefore unrealistic for a future-Earth for $S >$ 1.1. Similarly, 
photosynthetic activity for Earth biosystems has optimum temperatures
for gross primary productivity that peak between 293 and 308 K and experiences
sharp downward trends beyond 313 K \citep{bennett2021thermal}.

A major result of our work is that ozone could survive warm, damp (hence high HO$_\mathrm{X}$) conditions towards the IHZ due to mutual destruction of its catalytic sinks (HO$_\mathrm{X}$ by NO$_\mathrm{X}$). Previous works in literature regarding the study of steam atmosphere climates neglect O$_3$ in the atmosphere, owing to the increased presence of destructive HO$_\mathrm{X}$ species expected to appear in atmospheres dominated by water vapour \citep{kasting1988runaway,abe1988evolution,lebrun2013thermal,katyal2019evolution}. Similarly, with increasing proximity to the host star the increasing UV-B fluxes would be assumed to further contribute to the photolytic degradation of any ozone layer within the atmosphere. In our
work, the presence of a biosphere producing fluxes corresponding to the modern Earth (Table \ref{tab:lower_boundary_flux})
is found to maintain the O$_3$ layer centred on the atmospheric pressure range of 10$^0$--10$^1$ hPa despite the 
HO$_\mathrm{X}$ concentrations increasing by more than an order of magnitude due to enhanced water via ocean evaporation. As HO$_\mathrm{X}$ species react with NO$_\mathrm{X}$ species to form HO$_\mathrm{X}$-NO$_\mathrm{X}$ reservoir molecules such as HNO$_2$ and HNO$_3$, 
the weaker influence of NO$_\mathrm{X}$ as a sink for O$_3$ helps maintain the ozone layer for our simulated Earth-like planets in the IHZ. This could have important consequences for the atmospheric modelling of steam atmospheres, as
climate-only simulations could be underestimating the O$_3$ content of the middle atmosphere. As a consequence,
climate-chemistry coupled models of steam atmospheres that include a robust C-H-O-N chemistry scheme should be
investigated.

Theoretical spectra suggest that the O$_3$ spectral features produced in biotic simulations remain apparent in all 
instellation scenarios studied. The sharp fall in tropospheric O$_3$ is also 
shown to open a window to UV-C radiation at S $\geq$ 1.3, which is capable of stimulating ClO$_\mathrm{X}$ chemistry close
to the surface. As Cl and ClO are strong reducing agents, their impacts on atmospheric chemistry, especially ozone, close to the 
surface require further work in future photochemical modelling studies of planets close to or beyond the 
IHZ. We also show that despite the abiotic and biotic abundances differing substantially
for Earth-like planets (Fig. \ref{fig:BiovsAbio_Profiles}d), the resulting emission spectra signals are
difficult to differentiate from those originating from biologically inactive surfaces. To be a reliable biomarker,
extend observation times of greater than 5--10 days should be pursued for Earth-like planets within the habitable
zone with LIFE.

Although CH$_4$ is difficult to detect in the infrared emission spectra for planets producing Earth-like surface
fluxes, its presence can be inferred indirectly by looking at thermal emission features. For an `optimistic' 
aperture diameter of 3.5 m, LIFE would be able to detect the greenhouse heating of the atmosphere
induced by the H$_2$O produced by CH$_4$ oxidation for systems within 20 pc of our Solar System after
viewing periods of 5--10 days. The occurrence rate of Earth-like planets orbiting G-type stars with 
instellations exceeding 1.0, as estimated by \cite{Kopparapu2018}, has baseline estimates ranging from
0.21--0.67. A total of 141 G-type stars have been observed within 20 pc \citep{gaia2016,gaia2023} of our Solar System. 
Applying these Earth occurrence rates to this number of G-type stars, 29--94 Earth-like planets could be available for 
these measurements to target in future missions. 

Our modelling study produces substantially different results to the work of \cite{wolf2015evolution}, who
used a 3D climate model to assess the impacts on Earth's climate with increasing stellar instellation up to
a value of $S = 1.21$. They calculate global mean temperatures of approximately 360 K at $S = 1.21$ - at 
$S = 1.2$ in our study, surface temperatures only reach 318.93 K (Table \ref{tab:surftemps}) and 308.06 K in
our biotic and abiotic scenarios, respectively. These differences arise due differences in our water columns; \cite{wolf2015evolution} find a global mean column of $\sim$ 2500 kg m$^{-2}$ at $S = 1.21$, compared to $\sim$ 250 
kg m$^{-2}$ in our simulations at $S = 1.2$. When our model water column reaches $\sim$ 2125 kg m$^{-2}$ at $S = 1.5$
in our biotic scenarios, the greenhouse effect (calculated as the difference between the surface outbound longwave
radiation and the top of atmosphere outbound longwave radiation) are comparable to \cite{wolf2015evolution}.
These water column differences arise from the differences of the 3D and 1D models used. \cite{wolf2015evolution}
computes changing water vapour via 3D advection processes, turbulent mixing in the troposphere, and condensation
and evaporation. The results presented here, in the 1D approach, simply force the tropospheric water content
to the Manabe-Wetherald profile with a fixed surface relative humidity of 80\%, and calculate stratospheric
and mesospheric H$_2$O via vertical diffusion and photochemical calculations. Neglecting photochemical production
and loss of H$_2$O in the stratosphere and mesosphere also led climate studies such as \cite{wolf2015evolution}
to underestimate the atmospheric water content of modern Earth, which will lead to discrepancies appearing
between their work and studies such as ours that incorporate atmospheric photochemistry.

\section{Conclusions}\label{sec:conclusions}

Using 1D-TERRA to study the atmospheric composition of a planet hosting an Earth-like biosphere as solar instellation 
progressively increases has revealed the following key features:

\begin{itemize}
    \item For the biotic runs, the atmospheric O$_2$ content increases up to solar instellations of $S = 1.2$ ($\simeq$ 1603 W m$^{-2}$) due to, for example, catalytic recycling of HO$_\mathrm{X}$ into O$_2$. The increasing rate of H$_2$O and CO$_2$
    photolysis also enables additional abiotic O$_2$ to form faster than the increasing UV photolysis of O$_2$. Above $S = 1.2,$ 
    however, the ClO$_\mathrm{X}$  in the troposphere engages in photochemical loss cycles with O$_2$, which 
    causes abundances to start falling.     
    \item The stratospheric O$_3$ layer, centred between 10$^0$ and 10$^1$ hPa, is maintained above parts-per-million magnitudes up to
    and including runs with instellations of $S = 1.5$ ($\simeq$ 2042 W m$^{-2}$). This is counter to our initial hypothesis and the assumptions of prior climate models of water-rich atmospheres.
    With four orders of magnitude more H$_2$O than at $S = 1.0$, the increasing HO$_\mathrm{X}$ concentrations coupled
    with the increasing flux of UV (175.0--405.0 nm) radiation was expected to efficiently destroy the O$_3$ layer. Instead,
    the increasing concentration of O$_2$ from abiotic atmospheric chemistry and the suppression of ozone's NO$_\mathrm{X}$ sink from reactions with HO$_\mathrm{X}$ balance the increasing chemical and photolytic sinks.
    \item Biogenic CH$_4$ becomes unidentifiable in emission spectra after solar instellations
    of $S = 1.2$. The elevated HO$_\mathrm{X}$ and UV radiation fluxes greatly increase the 
    rate of photochemical loss, and beyond $S = 1.3$ the elevated ClO$_\mathrm{X}$ in the troposphere induces 
    chemical loss rates that push CH$_4$ to abundances below 0.1 pptv at atmospheric pressures below 10$^{2}$ hPa.
     \item UV-C radiation fluxes decrease with increasing solar instellation due to the increasing atmospheric H$_2$O content
    up to and including values of $S = 1.2$ ($\simeq$ 1633 W m$^{-2}$ total flux at the top of the atmosphere). Above $S = 1.2$,
    the tropospheric ozone abundances plummet due to enhanced ClO$_\mathrm{X}$ loss processes initiated by the 
    enhanced photolysis of outgassed HCl.
    \item The ozone band at 9.6 $\mu$m will require observation times greater than 120 hours for
    systems within 10 pc to reliably be designated a biosignature on an Earth-like planet if LIFE uses an aperture diameter of 2.0 m. Extending the aperture diameter to 3.5 m will enable the inference to be made for systems
    within 20 pc. 
    \item Thermal emission features between 14.0 and 16.0 $\mu$m in atmospheres with
    biological surface fluxes within 0.88 AU of their host stars are strong biomarkers, indicating the 
    presence of strong CH$_4$ modern-Earth-like surface emissions.
    Beyond 0.85 AU, the collapse of the abiotic scenarios' cold trap makes the abiotic
    and biotic thermal emission fluxes more difficult to distinguish.

\end{itemize}

\bibliographystyle{aa}    
\bibliography{main}

\begin{appendix}
\section{Additional volume mixing ratio profiles}

\begin{figure}
\centering
    \includegraphics[width=\hsize]{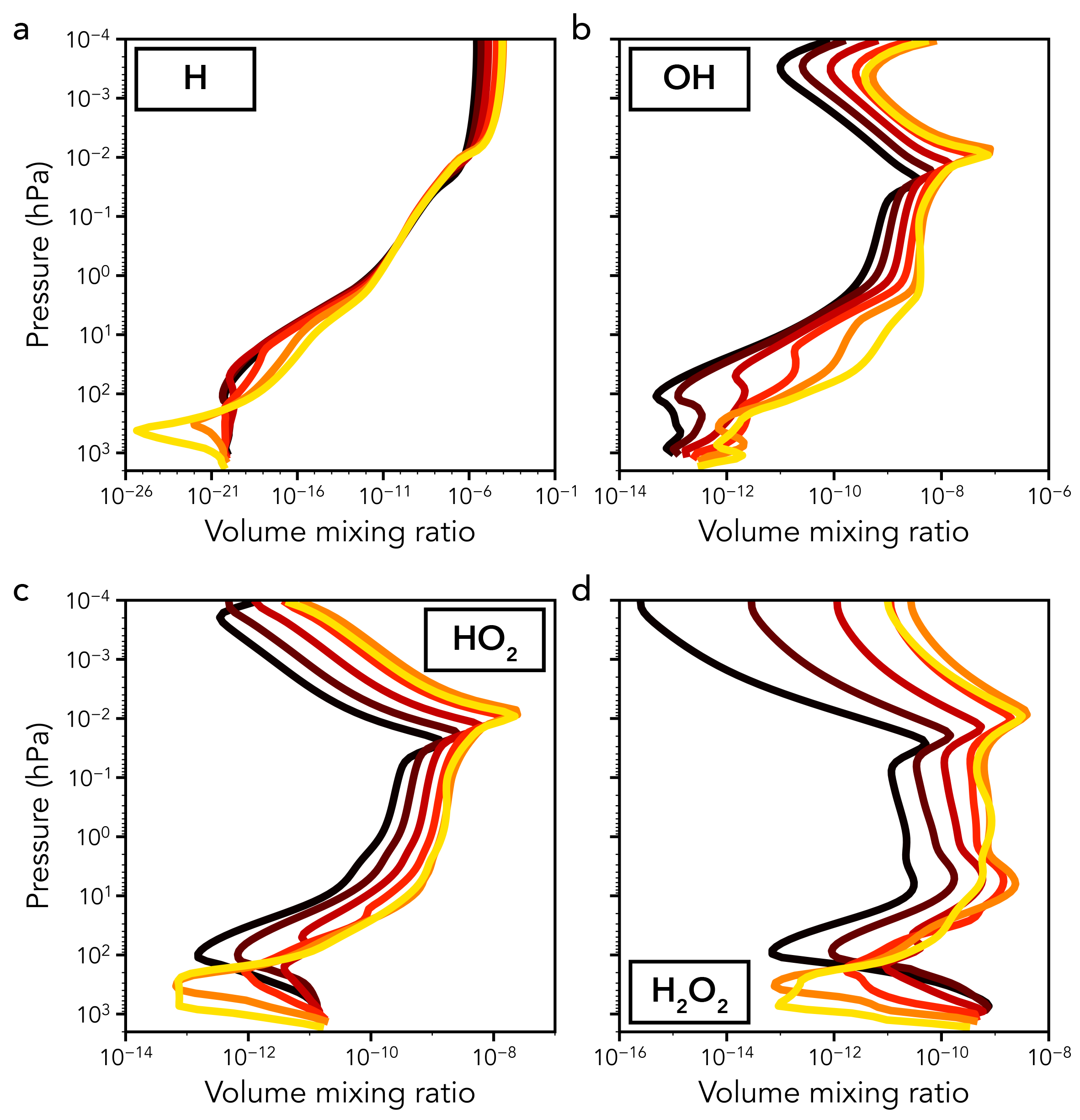}
    \caption{Individual VMRs of H (a), OH (b), and HO$_2$ (c) that summed together produce the profile of the
    HO$_\mathrm{X}$ chemical group (Figure~\ref{fig:chem_reactive}b), as well as H$_2$O$_2$ (d), a reservoir for HO$_\mathrm{X}$ species,
    for biotic model runs at solar constant 1.0, 1.1, 1.2, 1.3, 1.4, and 1.5 (see Figure~\ref{fig:climate} for the colour legend).}
\label{fig:HOX}
\end{figure}

\begin{figure}
\centering
    \includegraphics[width=\hsize]{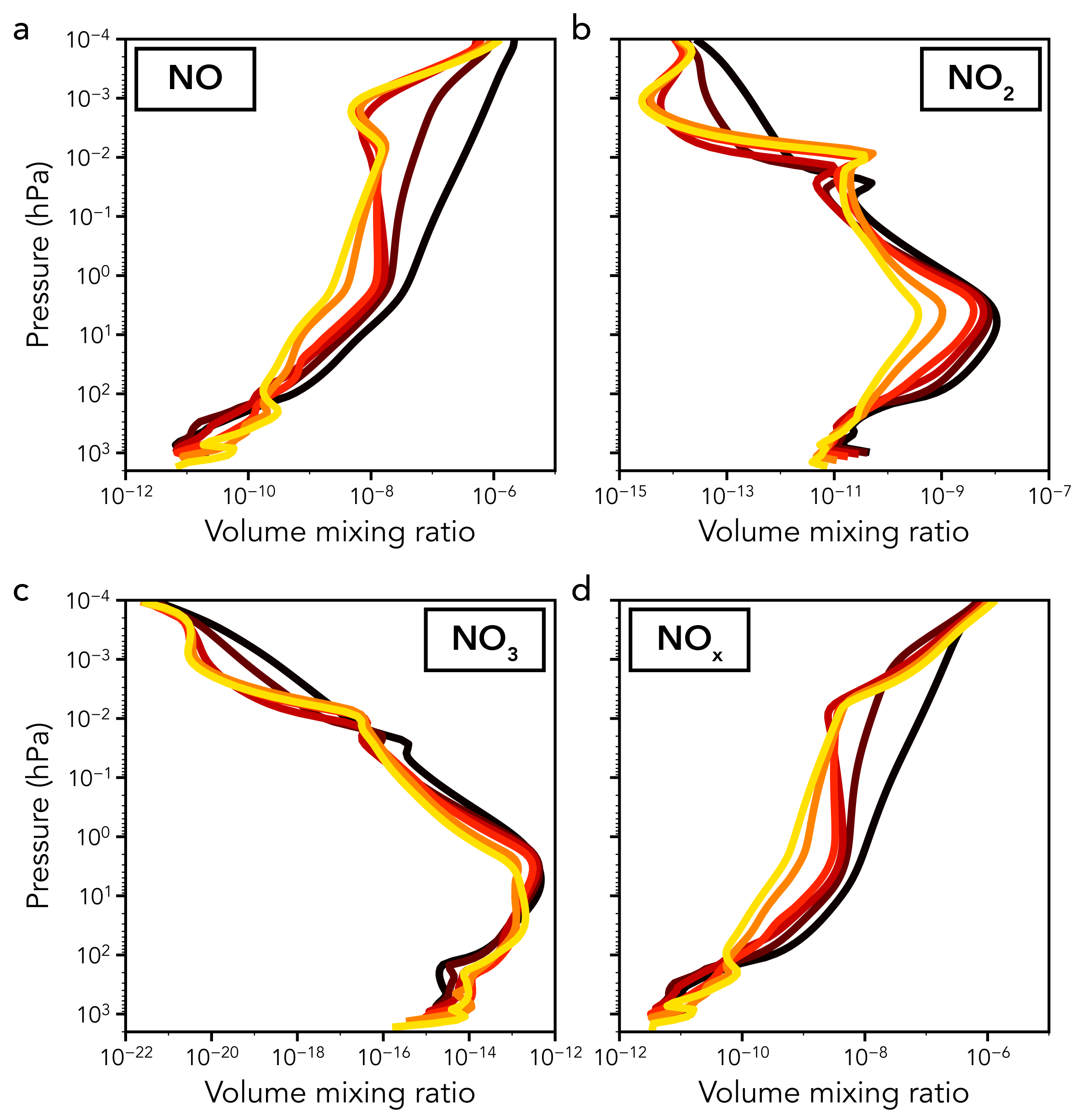}
    \caption{Individual VMRs of NO (a), NO$_2$ (b), and NO$_3$ (c) that summed together produce the profile of the
    NO$_\mathrm{X}$ chemical group (d; Figure~\ref{fig:chem_reactive}c) for biotic model runs at solar constant 1.0, 1.1, 1.2, 1.3, 1.4, and 1.5 (see Figure~\ref{fig:climate} for the colour legend).}
\label{fig:NOX}
\end{figure}

\begin{figure}
\centering
    \includegraphics[width=\hsize]{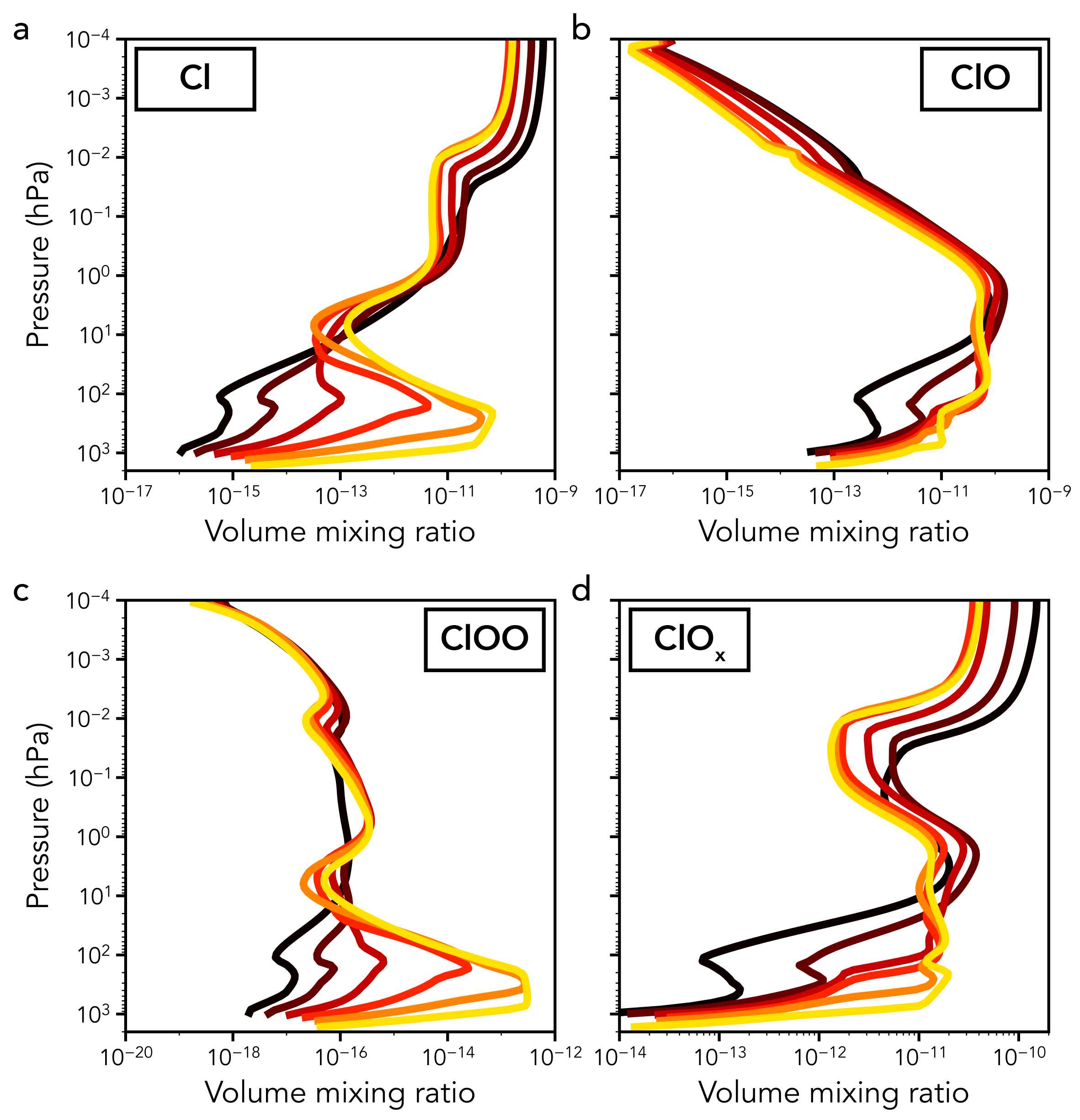}
    \caption{Individual VMRs of Cl (a), ClO (b), and ClOO (c) that summed together produce the profile of the
    ClO$_\mathrm{X}$ chemical group (d; Figure~\ref{fig:chem_others}c) for biotic model runs at solar constant 1.0, 1.1, 1.2, 1.3, 1.4, and 1.5 (see Figure~\ref{fig:climate} for the colour legend).}
\label{fig:ClOX}
\end{figure}

\section{Additional statistical significance study}

\begin{figure*}
\centering
    \includegraphics[width=\hsize]{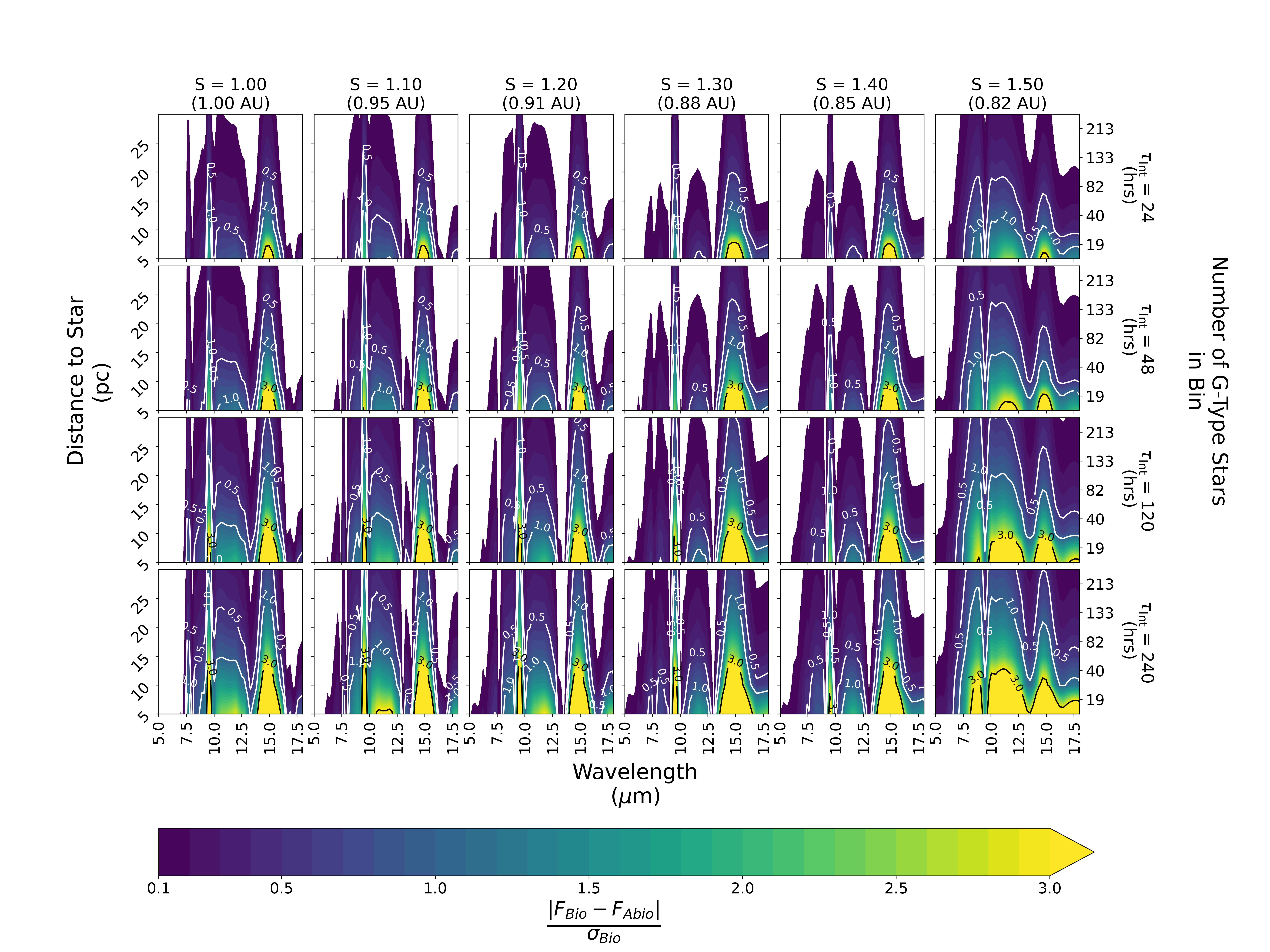}
    \caption{Same as Figure \ref{fig:LIFEsim_BiovsAbio}, but for LIFEsim's `optimistic' setting
    which assumes an aperture diameter of 3.5m, R = 50, quantum efficiency = 80\%, and throughput
    = 15\%.}
\label{fig:LIFEsim_BiovsAbio_Optimistic}
\end{figure*}

\end{appendix}

\end{document}